\documentclass[12pt]{article}
\usepackage{graphicx} % Required for inserting images
\usepackage[utf8]{inputenc}
\usepackage{geometry}
\usepackage{rotating}
\usepackage{floatrow}
\usepackage{rotating}
\usepackage{adjustbox}
\usepackage{amsfonts}
\usepackage{booktabs}
\usepackage{amssymb}
\usepackage{amsmath}
\usepackage{mathtools}
\usepackage[tableposition=top]{caption}
\usepackage{subcaption}
\usepackage{graphicx}
\usepackage{lscape}
\usepackage{bbm}
\usepackage{xcolor}
\usepackage{url}
\usepackage{footmisc}
\PassOptionsToPackage{hyphens}{url}
\usepackage[]{hyperref}

\newtheorem{result}{Result}

\newtheorem{lemma}{Lemma}

\newtheorem{proposition}{Proposition}
\newtheorem{remark}{Remark}

\newtheorem{assumption}{Assumption}
\usepackage{setspace}
\usepackage{footmisc}
\usepackage[round]{natbib}
\usepackage{float}
\floatstyle{plaintop}
\restylefloat{table}
\newenvironment{proof}[1][Proof]{\noindent \textbf{#1.} }{\  \rule{0.5em}{0.5em}}
\usepackage{enumerate}   
\usepackage{tikz}

\usepackage{accents}

\usepackage{chngcntr}
\usepackage{apptools}
\AtAppendix{\counterwithin{lemma}{section}}
\AtAppendix{\counterwithin{proposition}{section}}
\usepackage{kpfonts}
\usepackage[section]{placeins}

\title{Disagreement Spillovers\thanks{I thank Renee Bowen, Steven Brownstone, Gabriel Cañedo Riedel, Richard Carson, Julie Cullen, Simone Galperti, Nicola Gennaioli, Germán Gieczewski, Aram Grigoryan, Matias Iaryczower, Gleason Judd, John Londregan, Kris Ramsay, Denis Shishkin, Pietro Spini, Giorgio Saponaro, Joel Sobel, Rhet Smith, Guido Tabellini, Isabel Trevino, Manu Vespa, and participants to the Southern Economic Association for insightful comments and suggestions, and the Yankelovich Center of Social Policy %and the IHS (grant no. IHS017771) 
for financial support. The main RCT was registered as AEARCTR-0012729 \citep{bonomi2024aearctr12729}.}}
\author{Giampaolo Bonomi\thanks{Princeton University. bonomi@princeton.edu}}
\begin{document}
\maketitle

% =========================
% IMPROVED ABSTRACT
% =========================
% =========================
% REWRITTEN ABSTRACT (no change to structure; tighter logic)
% =========================

\begin{abstract}
Political messages increasingly bundle economic policy arguments with moral social policy stances. Using survey experiments with roughly 6,500 U.S.\ adults, I show that such bundling sharply weakens economic persuasion among respondents who disagree with the social stance: support falls by 13--20 percentage points relative to when the same economic message is sent alone, sometimes moving below pre-message levels. Bundling an aligned social stance does not increase persuasion. The main results are not driven by party cues, generalize across policy pairs, and are largely one-directional from social to economic issues, consistent with the predictions of a model of identity-based distancing.
\vspace{.35cm}

\end{abstract}

\newpage
\section{Introduction}\label{intro}

In U.S.\ politics, economic policy arguments are increasingly delivered alongside morally charged positions on social policy issues. Politicians and aligned media often discuss abortion, LGBTQ rights, or affirmative action in the same messages as taxation, trade, and redistribution.\footnote{All motivating trends are documented in Appendix \ref{motgraphs}.} Over the same period, disagreement on economic policy has become more tightly linked to disagreement on social and moral issues: voters are increasingly clustered into cultural factions whose views differ across both domains \citep{Gentzkow2016,BGT21,Desmet,Gethin}. This paper studies a simple micro-level channel connecting these developments: \emph{bundling} moral social-issue stances with economic recommendations changes how voters respond to economic arguments.

The empirical question is simple: holding an economic argument fixed, how does its persuasiveness change when it is paired with a moralized social policy stance from the same source? I find that moral disagreement generates large and robust cross-domain spillovers. When an economic recommendation is bundled with a moral stance that a respondent disagrees with, the respondent becomes substantially less persuaded by the economic recommendation. I call this effect \emph{disagreement spillovers}. In some cases, disagreement spillovers are strong enough to produce backlash, decreasing respondents' policy support relative to their prior, consistent with the idea that exposure to a culturally misaligned source can trigger adversarial reactions \citep[e.g.,][]{Norris_Inglehart_2019,Feddersen2022,Fukuyama2018}. In contrast, I find no evidence of corresponding \textit{agreement spillovers}: bundling an aligned moral stance does \textit{not} systematically increase persuasion relative to the same economic recommendation delivered without moral content.

These results have two broader implications. First, they extend the persuasion literature beyond the standard focus on explicit party labels or sender identities. Previous studies have shown that people react differently to information in the presence of partisanship cues \citep{Cohen2003}, and reviews highlight the general importance of cues and endorsements \citep{Druckman}. Recent work in economics similarly shows that attributing the same information to different political figures or political organizations changes how it is received \citep[e.g.,][]{BarberPope19,Afrouzi2024,ReyBiel}. Yet real political messages rarely contain a single cue or a single stance. Speeches, interviews, and ads routinely package multiple positions, and “who the speaker is” is often inferred from \emph{what else they say}. This paper isolates a within-message cross-domain interaction: a moral stance reshapes persuasion about an economic recommendation precisely because both are attributed to the same source.

Second, spillovers reshape the joint distribution of attitudes. Any heterogeneous persuasion response across moral groups can change how post-treatment economic attitudes map onto pre-treatment moral views, strengthening or weakening the observed association between social and economic positions. The asymmetry of my results makes the mechanism especially stark: bundling can strengthen moral--economic sorting even without increasing persuasion among aligned receivers, by selectively reducing persuasion (and sometimes inducing backlash) among misaligned receivers. Section \ref{scope} documents this implication directly by showing that the relationship between baseline moral views and post-treatment economic attitudes, initially weak, can substantially increase after bundled messages, generating correlated disagreement.

To identify these effects, I run pre-registered survey experiments on CloudResearch and Prolific with roughly 6{,}500 U.S.\ adults. In the main experiment, respondents read a message endorsing or opposing U.S.\ participation in the \textit{Comprehensive and Progressive Agreement for Trans-Pacific Partnership} (hereafter, CPTPP), either on its own or bundled with a pro-life or pro-choice abortion stance. Abortion views are elicited pre-treatment and CPTPP views post-treatment. A second experiment bundles a tax reform proposal with a stance supporting a ban on transgender adoptions.

Four empirical findings organize the paper. First, disagreement spillovers are large: relative to the same economic message without social content, misaligned bundling substantially reduces persuasion. Second, agreement spillovers are close to zero: aligned bundling does not meaningfully increase persuasion. Third, the effect is not simply moral priming and is not reducible to party cueing: the moral message alone does not appear to move economic attitudes, disagreement spillovers are not present when the moral and economic statements are attributed to different sources, but are robust to framing the source as non-partisan, controlling for party identity, or focusing on independents. Fourth, the pattern generalizes across policy pairs and is largely one-directional, from moral social issues to economic ones.

The results discipline interpretation. A natural benchmark is trust-based Bayesian updating: moral policy alignment may signal competence or shared interests, so receivers rationally weight economic recommendations differently depending on the moral stances of the source. Such models predict disagreement spillovers and can generate backlash under sufficiently strong perceived preference misalignment. However, in their most intuitive form, they also predict agreement spillovers to accompany disagreement spillovers. The data instead show strong negative spillovers with little positive spillover. Section \ref{interpretation} develops a conceptual framework that nests the trust benchmark and explains why this combination of facts is suggestive of a mechanism in which moral disagreement triggers identity-based distancing from the speaker.

The paper is organized as follows. Section \ref{literature} is a literature review. Section \ref{design} describes the experimental designs and estimands. Section \ref{results} presents the main evidence on spillovers and backlash. Section \ref{tests} reports diagnostic tests. Section \ref{scope} documents generality, one-directionality, and implications for the joint distribution of attitudes. Section \ref{interpretation} presents a conceptual framework and interpretation. Section \ref{conclusion} concludes.

\section{Related Literature}\label{literature}

The paper sits at the intersection of work on persuasion, identity politics, and the increasing multidimensionality of political conflict. The common theme across these literatures is that political messages affect beliefs and attitudes not only through their propositional content, but also through what they imply about the similarity between the source of political recommendation and the audience.

A first relevant strand studies \emph{source cues} and \emph{partisan alignment} in persuasion. Experimental evidence shows that support for the very same policy depends on whether it is endorsed by their party or the opposing party \citep{Cohen2003, Druckman}. Recent work in economics similarly documents that changing the political affiliation of the communicator---for instance, attributing an otherwise identical message to different political leaders or organizations---can substantially change persuasion \citep[e.g.,][]{BarberPope19,Afrouzi2024, ReyBiel}.

While these contributions typically vary an explicit politician or party label, my focus is on a feature of real political communication that is often taken as background: political messages routinely bundle multiple policy stances. Holding the economic recommendation fixed, I study how persuasion changes when the source also states a morally charged position on another issue in the same message. This design isolates a within-message cross-domain interaction that is distinct from (and can operate in addition to) standard cue effects. It is particularly relevant in settings where the receiver learns “what kind of person” the speaker is partly from what else the speaker says. In addition, by benchmarking comparisons to the case where economic messages are sent alone, I identify a systematic asymmetry between agreement and disagreement spillovers.

A second strand emphasizes the rise of \emph{identity politics} and the growing clustering of policy views across domains. Recent work argues that cultural identities and moral conflict have become increasingly central to political competition and voter behavior \citep{Norris_Inglehart_2019,Enke2020,GT23}. In parallel, evidence documents that U.S.\ voters have become more sorted across both social and economic dimensions, with shifting correlations among issue positions, demographic characteristics, and political identities \citep{Gentzkow2016,BGT21,Desmet,Gethin}. A complementary line of research documents systematic links between moral values and economic/political preferences \citep{Enke2020,Enke2022values,Enke2022,Enke2024}. My contribution to this conversation is to isolate a micro-level channel through which moral disagreement can \emph{causally} propagate to economic persuasion, generating correlated disagreement between voters.

A third relevant literature studies why elites link stances across issues in the first place. Models of platform choice and policy linkage show how parties may strategically bundle positions on social and economic issues to build coalitions or to structure political conflict \citep[e.g.,][]{Romer2007,BesleyPersson}. The mechanism studied here is complementary: even holding the economic argument itself fixed, bundling can affect persuasion by changing how receivers evaluate the speaker. In this sense, bundling is not only the strategic choice of what to promise to voters; it can also be a persuasion technology that opinion leaders can leverage to strategically reshape the joint distribution of voters' policy views.

Finally, the paper relates to research on \emph{trust in sources}. Economic agents may make assumptions on data generating processes that push them to rationally discount information depending on what it reveals about a source competence or bias; recent work develops microfoundations for disagreement about source accuracy and bias \citep{Gentzkow2024}.\footnote{Motivated reasoning provides a distinct route through which disagreement can shape updating \citep{Kunda1990TheCF}.} The trust-based benchmark in this paper formalizes disagreement and agreement spillovers as implications of how receivers map moral identity to trustworthiness (based on competence or bias); the identity-based interpretation instead builds on social identity theory and the idea that conflict and threat increase identity salience and outgroup differentiation \citep{TajfelWilkes,tajfel1979integrative,Oakes87,BranscombeThreat,Greenawaycruwys2018}. The asymmetry between disagreement and agreement effects is a tight constraint on mechanism and is naturally consistent with an identity-based distancing channel \citep{Parker13,BGT21,GT23}.

\section{Design}\label{design}

\subsection{Overview}\label{design:overview}

The design objective is to isolate a simple causal contrast: how does the persuasiveness of a fixed economic policy recommendation change when it is accompanied by morally charged social-policy content from the same source? I rely on three ingredients:

\begin{enumerate}[(i)]
\item \textbf{Randomized message content.} Respondents are randomly assigned to receive either no message, an economic recommendation alone, or an economic recommendation bundled with a social-policy stance.
\item \textbf{Pre-treatment measurement of moral views.} Because ``alignment'' is a property of the match between the respondent and the moral stance in the message, respondents report their baseline stance on the relevant social issue \emph{before} exposure. This pre-treatment measure is used only to classify respondents as aligned or misaligned with the moral stance they happened to receive; it is not used to determine treatment assignment.
\item \textbf{Post-treatment measurement of economic views.} Economic-policy attitudes are measured \emph{after} exposure, so differences across treatment arms capture persuasion effects rather than baseline differences.
\end{enumerate}

Two features of the design are important for interpretation. First, the economic argument is held fixed within each trade-treatment subsample: comparisons of bundled vs.\ unbundled messages keep the economic recommendation constant and vary only whether the source also expresses a moral stance. Second, because alignment is defined using pre-treatment views and assignment is random, aligned and misaligned receivers are comparable in expectation to the trade-only group; the aligned/misaligned effects can therefore be interpreted as causal effects of adding aligned/misaligned moral content to the same economic message.

\subsection{Main experiment: Abortion and trade}\label{design:trade}
The main CloudResearch study was preregistered in the AEA RCT Registry (AEARCTR-0012729) and fielded in July--August 2024.\footnote{The preregistration specified the core bundling design and its main estimating contrasts---comparisons across no-message, single-message, and bundled-message conditions used to test disagreement spillovers, agreement spillovers, and backlash---as well as the a priori hypothesis that these effects would be strongest from morally charged social issues to economic policy views; later diagnostics sharpen the interpretation of these preregistered patterns.} Respondents were randomized into seven arms: a no-message control, a pro-CPTPP message, an anti-CPTPP message, and four bundled-message arms that pair each trade recommendation with a pro-life or pro-choice abortion stance. The specific messages displayed to respondents and additional details about the experimental structure can be found in Appendix \ref{app:surveystructure}.

Let $Y_i\in\{0,1\}$ indicate whether respondent $i$ supports U.S.\ participation in the CPTPP. Let $T_i\in\{\text{Pro},\text{Anti},\varnothing\}$ denote the trade recommendation shown to $i$ (with $T_i=\varnothing$ in the no-message arm).\footnote{Outcomes are elicited on four-point support/oppose scales; the main tables report binary support indicators to facilitate interpretation. Ordered-response robustness is reported in Table \ref{ordered}.} Let $m_i\in\{0,1\}$ be $i$'s \emph{pre-treatment} abortion stance. In the bundled-message arms, let $s_i\in\{0,1\}$ be the abortion stance shown in the message; set $s_i=\varnothing$ when no abortion stance is shown. Define the alignment indicator $A_i\equiv \mathbbm{1}\{s_i=m_i\}$ (defined only when $s_i\neq\varnothing$) and the bundled-message indicator for trade recommendation $T$,
\[
B_i^T\equiv \mathbbm{1}\{T_i=T,\ s_i\neq\varnothing\}.
\]
For each $T\in\{\text{Pro},\text{Anti}\}$, define
\[
\text{Align}_i^{T}\equiv B_i^T A_i,\qquad
\text{Misalign}_i^{T}\equiv B_i^T(1-A_i).
\]
Respondents who receive the trade message $T$ without abortion content have $s_i=\varnothing$ and thus $\text{Align}_i^{T}=\text{Misalign}_i^{T}=0$; they form the natural within-message benchmark for spillovers.

\paragraph{Estimation.}
Within each trade-message subsample $T\in\{\text{Pro},\text{Anti}\}$, I estimate:
\begin{equation}\label{eq:spillover}
Y_i = \alpha_T + \tau_A^{T}\,\text{Align}_{i}^{T} + \tau_M^{T}\,\text{Misalign}_{i}^{T} + X_i'\gamma_T + \varepsilon_i,
\end{equation}
where $X_i$ is an optional vector of pre-treatment demographics. The omitted category is the trade-only message with the same trade recommendation $T$. I estimate \ref{eq:spillover} separately among respondents assigned to the pro-CPTPP (anti-CPTPP) recommendation arms, comparing bundled and unbundled within that recommendation. Note that $\tau_A^{T}$ and $\tau_M^{T}$ recover the average effect of bundling aligned and misaligned moral content, respectively, relative to sending the same economic recommendation alone.\footnote{Although alignment is defined using respondents' pre-treatment views, message assignment is independent of those views. With random assignment (and balanced arm sizes), the aligned group, misaligned group, and trade-only benchmark are all drawn from the same baseline distribution of pre-treatment moral stances; therefore the aligned/misaligned coefficients can be interpreted as differences in mean outcomes relative to the trade-only benchmark.}

\paragraph{Backlash.}
To test for backlash, I compare misaligned bundled arms to the no-message control (rather than the trade-only arm), asking whether misaligned bundling moves respondents past the no-message baseline in the direction opposite to the recommendation.

As an implementation detail, one bundled arm (Pro-Choice/Anti-CPTPP) began fielding a few days after the others. Appendix Table~\ref{spline} shows that adding flexible polynomial-spline controls for calendar time leaves the Table~\ref{mainresult} estimates essentially unchanged.

% ==========================================================
% SECTION 3: RESULTS
% ==========================================================
\section{Main Results}\label{results}

This section reports the main evidence from the abortion--trade experiment.

% ----------------------------
% INSERT TABLE mainresult HERE
% ----------------------------
% Paste your existing Table environment with \label{mainresult} here.

\subsection{Disagreement spillovers}\label{results:disagreement}
Table \ref{mainresult} provides strong support for disagreement spillovers. Panel A reports effects when the economic recommendation is \emph{pro}-CPTPP; Panel B when it is \emph{anti}-CPTPP. In each panel, the first pair of columns uses the trade-only message as the benchmark (a within-recommendation comparison: ``does adding moral content change persuasion relative to the same economic argument alone?''). The second pair of columns uses the no-message arm as the benchmark (a level comparison: ``does misaligned bundling move respondents past baseline, in the opposite direction of the recommendation?''). 

The disagreement spillovers in Table \ref{mainresult} are large in both sign and magnitude. In Panel A, bundling a \textit{pro}-CPTPP recommendation with a morally \emph{misaligned} abortion stance reduces CPTPP support by about 13 percentage points relative to the same pro-CPTPP message without abortion content. In Panel B, the same logic appears with the opposite sign: adding misaligned abortion content to an \emph{anti}-CPTPP recommendation increases CPTPP support by about 20 percentage points relative to the anti-CPTPP message alone. In other words, moral disagreement substantially weakens the persuasive force of the economic recommendation, regardless of whether the recommendation points “pro” or “anti.” This pattern is what makes the phenomenon a spillover: disagreement on the moral dimension interferes with updating on the economic dimension, even though the economic content is held fixed.

\begin{table}[htbp]
  \centering
  \caption{Spillovers and Backlash from Abortion to Trade}\label{mainresult}
\begin{adjustbox}{width=\textwidth}
    \begin{tabular}{p{13em}llllll}
    \toprule
    \multicolumn{1}{r}{} &       &       &       &       &       &  \\
    \multicolumn{1}{r}{} &       & \multicolumn{5}{c}{Differences in Share of} \\
    \multicolumn{1}{r}{} &       & \multicolumn{5}{c}{ Respondents Supporting the CPTPP} \\
    \multicolumn{1}{r}{} &       & \multicolumn{2}{c}{} &       & \multicolumn{2}{c}{} \\
    \midrule
    \multicolumn{1}{r}{} &       &       &       &       &       &  \\
    \multicolumn{1}{r}{} &       & \multicolumn{5}{c}{Panel A. Pro-CPTPP Message } \\
    \multicolumn{1}{r}{} &       &       &       &       &       &  \\
    \multicolumn{1}{r}{} &       & \multicolumn{2}{c}{Relative to} &       & \multicolumn{2}{c}{Relative to} \\
    \multicolumn{1}{r}{} &       & \multicolumn{2}{c}{Pro-CPTPP Only} &       & \multicolumn{2}{c}{No Message} \\
\cmidrule{3-7}    \multicolumn{1}{l}{Pro-CPTPP, Aligned Abortion (\(\tau_A^{\text{Pro}}\))} &       & \multicolumn{1}{c}{0.0244 } & \multicolumn{1}{c}{0.0244 } &       & \multicolumn{1}{c}{0.0438 } & \multicolumn{1}{c}{0.0467 } \\
    \multicolumn{1}{r}{} &       & \multicolumn{1}{c}{(0.0226)} & \multicolumn{1}{c}{(0.0221)} &       & \multicolumn{1}{c}{(0.0289)} & \multicolumn{1}{c}{(0.0288)} \\
    \multicolumn{1}{l}{Pro-CPTPP, Misaligned Abortion (\(\tau_M^{\text{Pro}}\))} &       & \multicolumn{1}{c}{-0.128***} & \multicolumn{1}{c}{-0.128***} &       & \multicolumn{1}{c}{-0.109***} & \multicolumn{1}{c}{-0.108***} \\
    \multicolumn{1}{r}{} &       & \multicolumn{1}{c}{(0.0271)} & \multicolumn{1}{c}{(0.0271)} &       & \multicolumn{1}{c}{(0.0326)} & \multicolumn{1}{c}{(0.0328)} \\
    \multicolumn{1}{r}{} &       &       &       &       &       &  \\
    \multicolumn{1}{l}{Observations} &       & \multicolumn{1}{c}{1254 } & \multicolumn{1}{c}{1251 } &       & \multicolumn{1}{c}{1050 } & \multicolumn{1}{c}{1047 } \\
    \multicolumn{1}{r}{} &       &       &       &       &       &  \\
    \midrule
    \multicolumn{1}{r}{} &       &       &       &       &       &  \\
    \multicolumn{1}{r}{} &       & \multicolumn{5}{c}{Panel B. Anti-CPTPP Message } \\
    \multicolumn{1}{r}{} &       &       &       &       &       &  \\
    \multicolumn{1}{r}{} &       & \multicolumn{2}{c}{Relative to} &       & \multicolumn{2}{c}{Relative to} \\
    \multicolumn{1}{r}{} &       & \multicolumn{2}{c}{Anti-CPTPP Only} &       & \multicolumn{2}{c}{No Message} \\
\cmidrule{3-7}    \multicolumn{1}{l}{Anti-CPTPP, Aligned Abortion (\(\tau_A^{\text{Anti}}\))} &       & \multicolumn{1}{c}{-0.00589} & \multicolumn{1}{c}{0.00234} &       & \multicolumn{1}{c}{-0.483***} & \multicolumn{1}{c}{-0.471***} \\
    \multicolumn{1}{r}{} &       & \multicolumn{1}{c}{(0.0327)} & \multicolumn{1}{c}{(0.0327)} &       & \multicolumn{1}{c}{(0.0333)} & \multicolumn{1}{c}{(0.0340)} \\
    \multicolumn{1}{l}{Anti-CPTPP, Misaligned Abortion (\(\tau_M^{\text{Anti}}\))} &       & \multicolumn{1}{c}{0.203***} & \multicolumn{1}{c}{0.200***} &       & \multicolumn{1}{c}{-0.275***} & \multicolumn{1}{c}{-0.273***} \\
    \multicolumn{1}{r}{} &       & \multicolumn{1}{c}{(0.0344)} & \multicolumn{1}{c}{(0.0335)} &       & \multicolumn{1}{c}{(0.0350)} & \multicolumn{1}{c}{(0.0346)} \\
    \multicolumn{1}{r}{} &       &       &       &       &       &  \\
    \multicolumn{1}{l}{Observations} &       & \multicolumn{1}{c}{1268} & \multicolumn{1}{c}{1268} &       & \multicolumn{1}{c}{1065} & \multicolumn{1}{c}{1064} \\
    \multicolumn{1}{r}{} &       &       &       &       &       &  \\
    \midrule
    \multicolumn{1}{r}{} &       &       &       &       &       &  \\
    \multicolumn{1}{l}{Demographic Controls} &       & \multicolumn{1}{c}{N} & \multicolumn{1}{c}{Y} &       & \multicolumn{1}{c}{N} & \multicolumn{1}{c}{Y} \\
    \multicolumn{1}{r}{} &       &       &       &       &       &  \\
    \midrule
    \midrule
    \multicolumn{7}{p{41em}}{\small Notes: Panel A reports the difference in the share of pro-CPTPP respondents between the groups who read the pro-CPTPP message bundled with an aligned and misaligned abortion message and the following two groups: (i) respondents who read only the pro-CPTPP message; and (ii) respondents in the No-Message condition. Panel B reports the difference in the share of pro-CPTPP respondents between the groups who read the anti-CPTPP message bundled with an aligned and misaligned abortion message and the following two groups: (i) respondents who read only the anti-CPTPP message; and (ii) respondents in the No-Message condition. Demographics include three age categories, and indicators for whether the respondent identifies as man, is white, is Christian,  is a Republican or a Democrat,  has a college degree, identifies as upper class or upper-middle class, and is unemployed.  Robust standard errors in parentheses. *** \(p<0.01\), ** \(p<0.05\), *\(p<0.1\).} \\
    \end{tabular}%
\end{adjustbox}
\end{table}

\begin{result}[Disagreement spillovers]\label{res:disagree}
Bundling an economic recommendation with a morally charged stance substantially reduces persuasion among respondents who disagree with the moral stance.
\end{result}

\subsection{No agreement spillovers}\label{results:agreement}
The aligned-bundling coefficients are small and statistically indistinguishable from zero in both panels. This is not a mechanical artifact of dichotomizing the outcome: Appendix Table \ref{ordered} shows the same qualitative pattern using the original four-point CPTPP measure in an ordered-response model.\footnote{Among respondents in the no-message control, answers on the four-point scale are concentrated at intermediate levels, with more than \(85\%\) of the respondents selecting ``somewhat support'' or ``somewhat oppose.''} The data therefore isolate a substantive asymmetry: moral disagreement reliably reduces persuasion, while moral agreement does not reliably raise persuasion beyond the economic message alone.

\begin{result}[Asymmetry]\label{res:asym}
Moral agreement does not increase persuasion: aligned bundling effects are close to zero, while misaligned bundling effects are large.
\end{result}

\subsection{Backlash}\label{results:backlash}
Table \ref{mainresult} also benchmarks bundled-message arms to the no-message control. Backlash is observed for pro-CPTPP recommendations (Panel A, ``Relative to No Message''): misaligned bundling reduces CPTPP support even below the no-message baseline, consistent with adversarial updating against a source perceived as morally unacceptable. For anti-CPTPP recommendations (Panel B), misaligned bundling still strongly attenuates persuasion relative to the anti-CPTPP-only benchmark, but does not produce the same clean reversal relative to no message. One interpretation is that the anti-CPTPP argument is, on average, highly persuasive even when the accompanying moral stance is objectionable (Panel B shows large negative movements relative to no message), whereas the pro-CPTPP argument is more fragile and can be overturned when the speaker is perceived as morally misaligned. The empirical takeaway is not that backlash is universal, but that bundling can sometimes move recipients past baseline in the opposite direction—an especially consequential form of spillover for political communication.

\begin{result}[Backlash]\label{res:backlash}
Misaligned bundling can generate backlash relative to the no-message baseline, particularly after pro-CPTPP recommendations.
\end{result}

\subsection{Robustness and heterogeneity}\label{results:robust}
Disagreement spillovers occur across the ideological spectrum. Table \ref{SpilBreak} decomposes the pooled spillover estimates by respondent abortion type and by the specific bundled message. The same qualitative pattern repeats across cells: for both pro-choice and pro-life respondents, persuasion is lower when the abortion stance in the bundle conflicts with their own view.

\vspace{0.5cm}
\begin{table}[H]
\begin{adjustbox}{width = \textwidth}
  \centering
  \caption{Spillover Effects Breakdown}\label{SpilBreak}
    \begin{tabular}{p{13.585em}llllll}
    \toprule
    \multicolumn{1}{r}{} &       &       &       &       &       &  \\
    \multicolumn{1}{r}{} &       & \multicolumn{5}{c}{Differences in Share of} \\
    \multicolumn{1}{r}{} &       & \multicolumn{5}{c}{ Respondents Supporting the CPTPP} \\
    \multicolumn{1}{r}{} &       & \multicolumn{5}{c}{(Relative to CPTPP Only)} \\
    \multicolumn{1}{r}{} &       &       &       &       &       &  \\
    \multicolumn{1}{r}{} &       & \multicolumn{2}{c}{Pro Life Respondents} &       & \multicolumn{2}{c}{Pro-Choice Respondents} \\
    \midrule
    \multicolumn{1}{r}{} &       &       &       &       &       &  \\
    \multicolumn{1}{l}{Pro-Life, Pro-CPTPP Message} &       & \multicolumn{1}{c}{0.0233} & \multicolumn{1}{c}{0.0409} &       & \multicolumn{1}{c}{-0.145***} & \multicolumn{1}{c}{-0.141***} \\
    \multicolumn{1}{r}{} &       & \multicolumn{1}{c}{(0.0436)} & \multicolumn{1}{c}{(0.0436)} &       & \multicolumn{1}{c}{(0.0257)} & \multicolumn{1}{c}{(0.0254)} \\
    \multicolumn{1}{r}{} &       & \multicolumn{1}{c}{[312]} & \multicolumn{1}{c}{[312]} &       & \multicolumn{1}{c}{[831]} & \multicolumn{1}{c}{[830]} \\
    \multicolumn{1}{r}{} &       &       &       &       &       &  \\
    \multicolumn{1}{l}{Pro-Life, Anti-CPTPP Message} &       & \multicolumn{1}{c}{0.0331} & \multicolumn{1}{c}{0.0504} &       & \multicolumn{1}{c}{0.224***} & \multicolumn{1}{c}{0.213***} \\
    \multicolumn{1}{r}{} &       & \multicolumn{1}{c}{(0.0587)} & \multicolumn{1}{c}{(0.0599)} &       & \multicolumn{1}{c}{(0.0391)} & \multicolumn{1}{c}{(0.0390)} \\
    \multicolumn{1}{r}{} &       & \multicolumn{1}{c}{[216]} & \multicolumn{1}{c}{[216]} &       & \multicolumn{1}{c}{[620]} & \multicolumn{1}{c}{[620]} \\
    \multicolumn{1}{r}{} &       &       &       &       &       &  \\
    \multicolumn{1}{l}{Pro-Choice, Pro-CPTPP Message} &       & \multicolumn{1}{c}{-0.122**} & \multicolumn{1}{c}{-0.122**} &       & \multicolumn{1}{c}{0.0147} & \multicolumn{1}{c}{0.00826} \\
    \multicolumn{1}{r}{} &       & \multicolumn{1}{c}{(0.0593)} & \multicolumn{1}{c}{(0.0606)} &       & \multicolumn{1}{c}{(0.0234)} & \multicolumn{1}{c}{(0.0227)} \\
    \multicolumn{1}{r}{} &       & \multicolumn{1}{c}{[231]} & \multicolumn{1}{c}{[230]} &       & \multicolumn{1}{c}{[606]} & \multicolumn{1}{c}{[604]} \\
    \multicolumn{1}{r}{} &       &       &       &       &       &  \\
    \multicolumn{1}{l}{Pro-Choice, Anti-CPTPP Message} &       & \multicolumn{1}{c}{0.124*} & \multicolumn{1}{c}{0.142**} &       & \multicolumn{1}{c}{-0.0163} & \multicolumn{1}{c}{-0.0138} \\
    \multicolumn{1}{r}{} &       & \multicolumn{1}{c}{(0.0655)} & \multicolumn{1}{c}{(0.0693)} &       & \multicolumn{1}{c}{(0.0384)} & \multicolumn{1}{c}{(0.0382)} \\
    \multicolumn{1}{r}{} &       & \multicolumn{1}{c}{[192]} & \multicolumn{1}{c}{[192]} &       & \multicolumn{1}{c}{[659]} & \multicolumn{1}{c}{[659]} \\
    \multicolumn{1}{r}{} &       &       &       &       &       &  \\
    \multicolumn{1}{l}{Demographic Controls} &       & \multicolumn{1}{c}{N} & \multicolumn{1}{c}{Y} &       & \multicolumn{1}{c}{N} & \multicolumn{1}{c}{Y} \\
    \multicolumn{1}{r}{} &       &       &       &       &       &  \\
    \midrule
    \midrule
    \multicolumn{7}{p{40em}}{Notes: The dependent variable is an indicator variable equal to 1 if the respondent supports the US participation in the CPTPP. Coefficients are differences relative to participants in the treatment arm where only the corresponding CPTPP message is sent. Demographic controls include three age categories, and indicators for whether the respondent identifies as a man, is white, is Christian,  is a Republican or a Democrat,  has a college degree, identifies as upper-class or upper-middle class, and is unemployed.  Robust standard errors are in parentheses. Sample size in square brackets. *** \(p<0.01\), ** \(p<0.05\), *\(p<0.1\)} \\
    \end{tabular}%
\end{adjustbox}
\end{table}%

Appendix Table \ref{AggrBreak} similarly shows that backlash patterns are not confined to a single moral subgroup.\footnote{To increase statistical power in the estimates of Tables \ref{SpilBreak} and \ref{AggrBreak},  I included in the corresponding estimation samples responses from a small pilot of the survey experiment conducted on Cloud Research the day before starting the main experiment. The small pilot contained a Pro-CPTPP arm, a Pro-Life, pro-CPTPP arm, and a No-message arm, all identical to those in the main experiment. Results are almost entirely unchanged (in terms of significance and effect sizes) if this sample is omitted.} Together, these decompositions indicate that the pooled disagreement spillovers in Table \ref{mainresult} are not an artifact of a single idiosyncratic pairing; rather, misalignment consistently undermines persuasion across message variants.

The next section uses additional experimental variation to distinguish between competing explanations. In particular, it shows that the effect survives controls for party cues, is not generated by moral content in isolation (a priming-only account), and that it largely disappears when the moral and economic statements are attributed to different sources. This common-source requirement is a strong constraint on mechanism: it suggests that what matters is not merely that moral conflict becomes salient, but that the conflict attaches to the \emph{speaker} of the economic recommendation.

% ==========================================================
% SECTION 4: DIAGNOSTICS
% ========================================================

\section{Beyond Party Cues and Moral Salience}\label{tests}

The main results could reflect multiple channels. This section uses diagnostic designs to narrow the set of plausible explanations.

\paragraph{Beyond party-cue interpretations.}\label{tests:party}
Because abortion stances correlate with party identity, one might interpret the moral stance mainly as a partisan cue. Three findings argue against this interpretation. Disagreement spillovers remain large among independents (Table \ref{independents}, Appendix), they are robust to party-cue controls (Table \ref{partycues}, Appendix), and they persist when the speaker is explicitly framed as non-partisan (Table \ref{nopty}, below).

\begin{table}[h!]
  \centering
  \caption{Spillovers from Abortion to Trade: Non-Partisan Source}\label{nopty}
    \begin{tabular}{p{14.085em}llll}
    \toprule
    \multicolumn{1}{r}{} &       &       &       &  \\
    \multicolumn{1}{r}{} &       & \multicolumn{3}{c}{Change in Share of} \\
    \multicolumn{1}{r}{} &       & \multicolumn{3}{c}{ Respondents Supporting the CPTPP} \\
    \multicolumn{1}{r}{} &       & \multicolumn{3}{c}{(Relative to Pro-CPTPP Only)} \\
    \multicolumn{1}{r}{} &       &       &       &  \\
    \multicolumn{1}{r}{} &       & Pro-Choice Respondents &       & Pro-Life Respondents \\
    \midrule
    \multicolumn{1}{r}{} &       &       &       &  \\
    \multicolumn{1}{l}{Pro-Life, Pro-CPTPP Message (\(\beta\))} &       & \multicolumn{1}{c}{-0.113***} &       & \multicolumn{1}{c}{0.0605 } \\
    \multicolumn{1}{r}{} &       & \multicolumn{1}{c}{(0.0359)} &       & \multicolumn{1}{c}{(0.0769)} \\
    \multicolumn{1}{r}{} &       &       &       &  \\
    \multicolumn{1}{l}{Separate Messages (\(\gamma\))} &       & \multicolumn{1}{c}{0.0066 } &       & \multicolumn{1}{c}{-0.0514} \\
    \multicolumn{1}{r}{} &       & \multicolumn{1}{c}{(0.0360)} &       & \multicolumn{1}{c}{(0.0741)} \\
    \multicolumn{1}{r}{} &       &       &       &  \\
    \multicolumn{1}{l}{Observations} &       & \multicolumn{1}{c}{1118} &       & \multicolumn{1}{c}{374} \\
    \multicolumn{1}{r}{} &       &       &       &  \\
    \midrule
    \midrule
    \multicolumn{5}{p{38.75em}}{\small Notes: The dependent variable is an indicator variable equal to 1 if the respondent supports the US membership in the CPTPP. Coefficients are differences relative to participants in the treatment arm where only the Pro-CPTPP message is sent. In all treatment arms, messages are described as coming from non-partisan sources. Estimation is carried out with a FE model, and standard errors are clustered at the individual level. *** \(p<0.01\), ** \(p<0.05\), *\(p<0.1\).} \\
    \end{tabular}%
\end{table}%

Table \ref{nopty} is based on a follow-up study, which re-interviews a subset of respondents and measures CPTPP support both before and after exposure to a second-round message. Let $Y_{it}\in\{0,1\}$ be an indicator for supporting CPTPP for individual $i$ at wave $t\in\{\text{pre},\text{post}\}$. Let $\text{Post}_t$ indicate the post-treatment wave, and let $\text{Bundled}_i$ and $\text{Separate}_i$ indicate assignment to the bundled (pro-life + pro-CPTPP from the same non-partisan speaker) and separate-speakers conditions, respectively (the omitted category is pro-CPTPP-only). In all conditions, the speaker is framed as \textit{not affiliated with any political party}. I estimate:
\begin{equation}\label{eq:followup}
Y_{it}=\alpha_i+\lambda\,\text{Post}_t+\beta\text{Bundled}_i\times \text{Post}_t+\gamma\text{Separate}_i\times \text{Post}_t+u_{it},
\end{equation}
by baseline abortion type, with standard errors clustered at the individual level. Equivalently, this is the first-difference regression $\Delta Y_i=\lambda +\beta\,\text{Bundled}_i+\gamma\,\text{Separate}_i+\Delta u_i$.

\paragraph{Spillovers require a common source}\label{tests:samesource}
The strongest diagnostic, reported in Table \ref{nopty}, holds content fixed but varies whether the moral and economic statements are attributed to the same source. When respondents read the same two passages but are told they come from different speakers (``Separate Messages''), spillovers disappear: the coefficient \(\gamma\), capturing the economic persuasion effect of the additional pro-life stance, has a zero point estimate among pro-choice respondents.

\begin{result}[Common-source]\label{res:samesource}
Disagreement spillovers are largely deactivated when the moral and economic statements are attributed to different sources, holding content fixed.
\end{result}

\paragraph{Priming placebo: not just salience.}\label{tests:priming}
Table \ref{priming} in the Appendix shows that presenting the pro-life abortion message without any trade content does not meaningfully affect CPTPP support relative to no message. This suggests the main effects are not driven by moral salience alone; they arise when moral content is bundled with an economic recommendation from the same source.\footnote{On top of the seven main arms described in Section \ref{design:trade}, the Cloud Research experiment included additional treatment arms to test for priming effects (Table \ref{priming}) and persuasion spillovers from trade to abortion policy (Table \ref{ReverseSpillovers}, Panel A).}

% ==========================================================
% SECTION 5: SCOPE AND DIRECTION
% ==========================================================
\section{Scope, Direction, and Implications for Disagreement}\label{scope}

This section extends the main abortion--trade findings in three ways. First, it shows that disagreement spillovers are not specific to one policy pair, one survey platform, or one particular moral stance: they also arise when a tax reform proposal is bundled with a stance supporting a ban on transgender adoptions. Second, it clarifies directionality: while moral disagreement spills over into economic persuasion, I find little evidence of spillovers running from economic issues back into moral social-policy views. Third, it explains why these patterns matter for multidimensional polarization: bundled communication can shift the mapping from moral identities to economic attitudes, strengthening or weakening the observed association between social and economic positions in the electorate.

\begin{table}[h!]
\begin{adjustbox}{width=\textwidth}
  \centering
  \caption{Spillovers from Transgender Rights to Taxation}\label{transtax}
    \begin{tabular}{p{11.915em}llllll}
    \toprule
    \multicolumn{1}{r}{} &       &       &       &       &       &  \\
    \multicolumn{1}{r}{} &       & \multicolumn{5}{c}{Differences in Share of} \\
    \multicolumn{1}{r}{} &       & \multicolumn{5}{c}{ Respondents Supporting the Tax Reform} \\
    \multicolumn{1}{r}{} &       & \multicolumn{5}{c}{(Relative to Pro-Tax-Reform Only)} \\
    \multicolumn{1}{r}{} &       &       &       &       &       &  \\
    \multicolumn{1}{r}{} &       & \multicolumn{2}{c}{Anti-Ban Respondents} &       & \multicolumn{2}{c}{Pro-Ban Respondents} \\
    \midrule
    \multicolumn{1}{r}{} &       &       &       &       &       &  \\
    \multicolumn{1}{l}{Pro-Ban, Pro-Tax-Reform Message} &       & \multicolumn{1}{c}{-0.0806**} & \multicolumn{1}{c}{-0.0795**} &       & \multicolumn{1}{c}{-0.0188} & \multicolumn{1}{c}{-0.0155} \\
    \multicolumn{1}{l}{} &       & \multicolumn{1}{c}{(0.0396)} & \multicolumn{1}{c}{(0.0393)} &       & \multicolumn{1}{c}{(0.0464)} & \multicolumn{1}{c}{(0.0466)} \\
    \multicolumn{1}{l}{} &       &       &       &       &       &  \\
    \multicolumn{1}{l}{Observations} &       & \multicolumn{1}{c}{604} & \multicolumn{1}{c}{603} &       & \multicolumn{1}{c}{376} & \multicolumn{1}{c}{374} \\
    \multicolumn{1}{l}{Demographic Controls} &       & \multicolumn{1}{c}{N} & \multicolumn{1}{c}{Y} &       & \multicolumn{1}{c}{N} & \multicolumn{1}{c}{Y} \\
    \multicolumn{1}{r}{} &       &       &       &       &       &  \\
    \midrule
    \midrule
    \multicolumn{7}{p{38.58em}}{\small Notes: The dependent variable is an indicator variable equal to 1 if the respondent supports the tax reform proposal. Coefficients are differences relative to participants in the treatment arm where only the taxation message is sent. Demographic controls include three age categories, and indicators for whether the respondent identifies as man, is white, is Christian, is a Republican or a Democrat,  has a college degree, identifies as upper-class or upper-middle class, and is unemployed.  Robust standard errors in parentheses.  *** \(p<0.01\), ** \(p<0.05\), *\(p<0.1\).} \\
    \end{tabular}%
\end{adjustbox}
\end{table}%

\subsection{Generality: transgender adoption and taxation}\label{scope:general}
A natural question is whether disagreement spillovers are a peculiarity of the abortion-trade policy pair. Table \ref{transtax} addresses this concern by testing a different moral--economic bundle with a second, analogous, experiment run on Prolific in August 2024. Respondents read a proposal to cut income taxes and increase a VAT, either alone or bundled with a message endorsing a ban on transgender adoptions.

Two features of Table \ref{transtax} are worth emphasizing. First, disagreement spillovers replicate: respondents who oppose the adoption ban become about 8 percentage points less likely to support the tax reform when the reform is bundled with the ban endorsement. Second, aligned bundling again produces little additional persuasion: pro-ban respondents are not significantly more persuaded by the tax reform when it comes from a pro-ban source. The replication therefore supports an interpretation in which morally charged disagreement is a systematic ``interference'' channel in persuasion, rather than an artifact of abortion rhetoric or of trade policy in particular. The fact that similar spillovers arise for different moral issues suggests that what matters is not the informational content of the specific social issue, but the way morally loaded stances change the receiver's willingness to ``follow'' the same source across domains.

\subsection{Directionality: no spillovers from economics to moral issues}\label{scope:reverse}
If bundling reveals something about a source and induces broad ideological updating, one might expect the economic stance to spill over into social policy as well. To test for spillover from the economic to the cultural domain, I included treatment arms that compare the persuasive power of a moral social policy argument (on abortion or transgender rights) when displayed alone vs with an economic policy stance (on the trade treaty or the tax reform).\footnote{In these arms, economic policy views are measured before treatment, while moral social policy views are post-treatment outcomes.} I find little evidence of these reverse spillovers (Table \ref{ReverseSpillovers}). Across both settings, bundling does not meaningfully shift average moral-policy views as a function of baseline economic positions. In other words, while moral disagreement affects economic persuasion, exposure to economic content does not comparably move moral views.

This directional asymmetry is informative because it helps distinguish mechanisms. If the main channel were simply that bundling primes ideological thinking or raises general political affect, then the economic message should plausibly move moral attitudes too. Instead, the evidence is more consistent with an ``anchor'' view of moral issues: moral stances are comparatively stable, well-formed, identity-linked positions, while economic attitudes on complex policies are more malleable. Hence, morally charged issues function as \emph{triggers} and the economic issues studied here function as \emph{targets}. This is also consistent with evidence that moral values are tightly connected to identity and political sorting \citep{Enke2020,Enke2022values,Enke2022,Enke2024,BGT21}.

\begin{table}[H]
\begin{adjustbox}{width=\textwidth}
  \centering
  \caption{Spillovers from Economic Policy to Social Policy}\label{ReverseSpillovers}
    \begin{tabular}{p{16.915em}llllll}
    \toprule
    \multicolumn{1}{r}{} &       &       &       &       &       &  \\
    \multicolumn{1}{r}{} &       & \multicolumn{5}{c}{Panel A. Differences in Share of} \\
    \multicolumn{1}{r}{} &       & \multicolumn{5}{c}{ Pro-Life Respondents } \\
    \multicolumn{1}{r}{} &       & \multicolumn{5}{c}{(Relative to Pro-Life Only)} \\
    \multicolumn{1}{r}{} &       &       &       &       &       &  \\
    \multicolumn{1}{r}{} &       & \multicolumn{2}{c}{Anti-CPTPP} &       & \multicolumn{2}{c}{Pro-CPTPP} \\
    \multicolumn{1}{r}{} &       & \multicolumn{2}{c}{Respondents} &       & \multicolumn{2}{c}{Respondents} \\
\cmidrule{3-7}    \multicolumn{1}{l}{Pro-CPTPP, Pro-Life Message} &       & \multicolumn{1}{c}{-0.012} & \multicolumn{1}{c}{-0.058} &       & \multicolumn{1}{c}{0.0538*} & \multicolumn{1}{c}{0.0498*} \\
    \multicolumn{1}{r}{} &       & \multicolumn{1}{c}{(0.0894)} & \multicolumn{1}{c}{(0.0859)} &       & \multicolumn{1}{c}{(0.0313)} & \multicolumn{1}{c}{(0.0258)} \\
    \multicolumn{1}{r}{} &       &       &       &       &       &  \\
    \multicolumn{1}{l}{Observations} &       & \multicolumn{1}{c}{131 } & \multicolumn{1}{c}{131 } &       & \multicolumn{1}{c}{713 } & \multicolumn{1}{c}{711 } \\
    \multicolumn{1}{r}{} &       &       &       &       &       &  \\
    \midrule
    \multicolumn{1}{r}{} &       &       &       &       &       &  \\
    \multicolumn{1}{r}{} &       & \multicolumn{5}{c}{Panel B. Differences in Share of} \\
    \multicolumn{1}{r}{} &       & \multicolumn{5}{c}{Pro-Adoption-Ban Respondents} \\
    \multicolumn{1}{r}{} &       & \multicolumn{5}{c}{(Relative to Pro-Ban Only)} \\
    \multicolumn{1}{r}{} &       &       &       &       &       &  \\
    \multicolumn{1}{r}{} &       & \multicolumn{2}{c}{Anti-Tax-Reform} &       & \multicolumn{2}{c}{Pro-Tax-Reform} \\
\multicolumn{1}{r}{} &       & \multicolumn{2}{c}{  Respondents} &       & \multicolumn{2}{c}{ Respondents } \\
\cmidrule{3-7}    \multicolumn{1}{l}{Pro-Tax-Reform, Pro-Ban Message} &       & \multicolumn{1}{c}{-0.0356} & \multicolumn{1}{c}{-0.0187} &       & \multicolumn{1}{c}{-0.0183} & \multicolumn{1}{c}{-0.0211} \\
    \multicolumn{1}{r}{} &       & \multicolumn{1}{c}{(0.0445)} & \multicolumn{1}{c}{(0.0431)} &       & \multicolumn{1}{c}{(0.0371)} & \multicolumn{1}{c}{(0.0350)} \\
    \multicolumn{1}{r}{} &       &       &       &       &       &  \\
    \multicolumn{1}{l}{Observations} &       & \multicolumn{1}{c}{389} & \multicolumn{1}{c}{388} &       & \multicolumn{1}{c}{584} & \multicolumn{1}{c}{581} \\
    \multicolumn{1}{r}{} &       &       &       &       &       &  \\
    \midrule
    \multicolumn{1}{r}{} &       &       &       &       &       &  \\
    \multicolumn{1}{l}{Demographic Controls} &       & \multicolumn{1}{c}{N} & \multicolumn{1}{c}{Y} &       & \multicolumn{1}{c}{N} & \multicolumn{1}{c}{Y} \\
    \multicolumn{1}{r}{} &       &       &       &       &       &  \\
    \midrule
    \midrule
    \multicolumn{7}{p{35.83em}}{\small Notes: In Panel A, the dependent variable is an indicator variable equal to 1 if the respondent is Pro-Life. Coefficients are differences relative to participants in the treatment arm where only the Pro-Life message is sent. In Panel B, the dependent variable is an indicator variable equal to 1 if the respondent is Pro-Ban. Coefficients are differences relative to participants in the treatment arm where only the Pro-Ban message is sent. Demographic controls include three age categories and indicators for whether the respondent identifies as a man, is white, is Christian, is a Republican or a Democrat,  has a college degree, identifies as upper-class or upper-middle class, and is unemployed.  Robust standard errors in parentheses. *** \(p<0.01\), ** \(p<0.05\), * \(p<0.1\).} \\
    \end{tabular}%
\end{adjustbox}
\end{table}%

\subsection{Correlated Disagreement}\label{scope:corr}
Finally, the results have implications for multidimensional polarization. Existing work documents that policy views in the U.S.\ have become more clustered across domains \citep{Gentzkow2016,BGT21,Desmet}. The experiments provide a micro-level mechanism for how such clustering can be reinforced (or weakened) by communication.

A practical implication is that the persuasion constraint imposed by moral disagreement is asymmetric: moral conflict can undermine economic persuasion even when the economic argument is held fixed, but economic disagreement is less likely to soften or harden moral stances. Hence, elites may repeatedly bundle moral conflict into broader messages without necessarily shifting the underlying moral divide.

\begin{figure}[H]
\centering
\includegraphics[width=.85\linewidth]{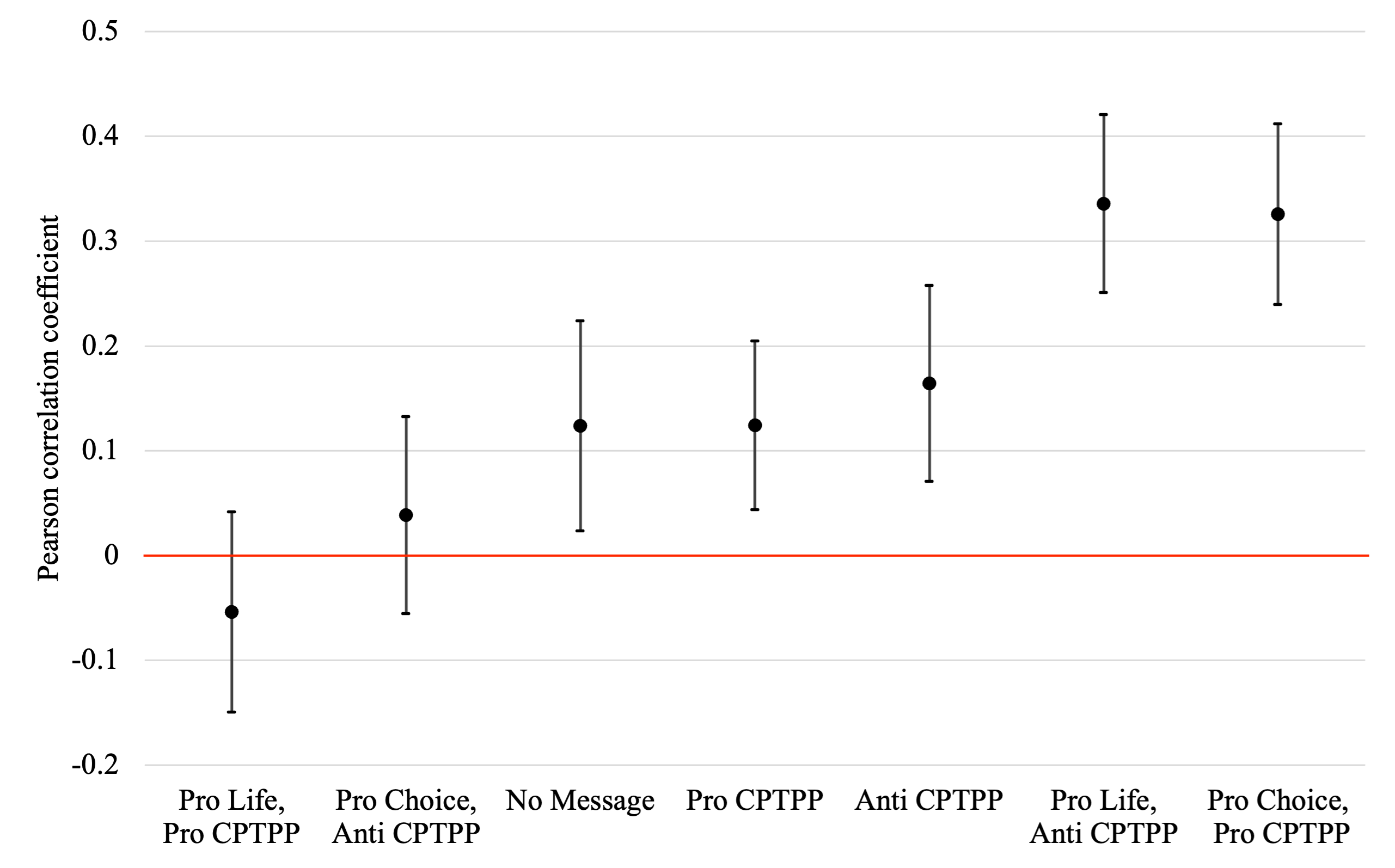}
    \caption{Correlation between Pro-Choice and Pro-CPTPP views by Treatment}\label{correxp2}
    \floatfoot{Notes. For each of the treatment conditions, the figure plots the Pearson correlation between support for the CPTPP and pro-choice views in the corresponding treatment arm. Support for the CPTPP and pro-choice views are measured by indicator variables equal to \(1\), respectively, if the respondent supports the US membership in the CPTPP and if the respondent is against anti-abortion laws. \(95\%\) confidence intervals reported. Source: own survey data.}
\end{figure}

The mechanism has an important implication for the \emph{joint distribution} of attitudes. Because baseline moral views are measured pre-treatment, the experiment can directly compare how the relationship between moral identity and economic attitudes differs across message arms. Any setting in which respondents with different moral views respond differently to the same economic recommendation will mechanically shift the mapping from $m_i$ to $Y_i$ and therefore can change the observed correlation between moral and economic views. Positive agreement spillovers are not required for this: it is enough that misaligned respondents discount the message, update less, or react against it. Conversely, if a bundle produces negative spillovers in a group that would otherwise update toward the recommendation, it can either strengthen or weaken sorting depending on how the recommended economic stance relates to the groups' baseline economic tendencies. 

Figure \ref{correxp2} illustrates this logic by plotting the correlation between pro-choice views and CPTPP support by treatment arm. In the no-message arm, the correlation is positive but modest. Bundled arms that effectively ``select'' persuasion into one moral group—by depressing persuasion among misaligned respondents—tend to increase the correlation, while bundles that counteract baseline sorting can attenuate it. The figure suggests that bundling can shift the joint distribution in predictable ways through asymmetric persuasion responses.

Taken together, the generality across policy pairs, the largely one-directional spillover pattern, and the joint-distribution evidence point to a broad takeaway: when moral conflict is attached to the speaker of an economic recommendation, it can operate as a powerful constraint on economic persuasion and, through heterogeneous reactions, contribute to the clustering of economic attitudes along moral lines.

\section{Conceptual Framework and Interpretation}\label{interpretation}

The experiments establish three empirical regularities: (i) moral disagreement reduces persuasion on an economic recommendation delivered by the same source; (ii) moral agreement produces little or no additional persuasion; and (iii) the disagreement effect largely disappears when the moral and economic statements are attributed to different sources. This section presents a conceptual framework that clarifies what classes of models predict, why the observed asymmetry is informative, and why the common-source requirement matters.

\subsection{A trust-based Bayesian benchmark}\label{interpretation:trust}

Consider a single economic policy choice, coded as $1$ (support) or $0$ (oppose).
A receiver $R$ is uncertain whether supporting the policy is welfare-improving for her.
Let $\omega^R\in\{0,1\}$ denote the welfare-relevant state, where $\omega^R=1$ means the policy is ``good for $R$'' (so a fully informed receiver would want to support it) and $\omega^R=0$ means it is ``bad for $R$.''%
\footnote{The experiments measure stated support for the policy. The trust benchmark is stated in terms of posterior beliefs about $\omega^R$; to map beliefs to measured support it is enough to assume that (in expectation) respondents are more likely to support the policy when they assign higher posterior probability to $\omega^R=1$.}

A source $S$ takes an economic stance $y^S\in\{0,1\}$, where $y^S=1$ denotes a recommendation to support the policy (and $y^S=0$ denotes a recommendation to oppose it). The receiver may also observe the source's moral/cultural type $\theta\in\{A,M\}$: aligned ($A$) or misaligned ($M$) with the receiver on a salient moral issue. The receiver's \emph{mental model} is a joint distribution $\mathbb{P}$ over $(\omega^R,y^S,\theta)$ with full support.

In this benchmark, moral alignment matters because it shapes what the receiver infers from the same economic recommendation. One possibility is a \emph{competence} channel: the receiver treats aligned sources as more informed about the economic state, so the stance $y^S$ is read as stronger evidence about $\omega^R$ when $\theta=A$ than when $\theta=M$. A second possibility is a \emph{preference} channel: the receiver treats moral alignment as a signal of shared goals, so $y^S$ is interpreted as more likely to be in the receiver's interest when $\theta=A$ than when $\theta=M$. Appendix \ref{trustbased} formalizes both channels and shows how they generate spillovers and, under sufficiently strong perceived preference misalignment, backlash.

Either way, the implication for the empirical design is that revealing $\theta=M$ can reduce the persuasiveness of a fixed economic recommendation relative to the ``prior-based'' case where the receiver observes $y^S$ but not $\theta$. The same logic also clarifies why the same-source diagnostic is informative: if the moral statement is attributed to a different speaker, it stops being evidence about the trustworthiness of the source of $y^S$, so trust-based spillovers should attenuate under separation.

\subsection{Why the asymmetry is informative}\label{interpretation:asym}

The trust benchmark also highlights why the absence of agreement spillovers is not innocuous in the simplest case. Fix an economic recommendation $y^S=1$ and define the receiver's posterior belief that the policy is good as
\begin{align*}
p(A) &\equiv \mathbb{P}(\omega^R=1\mid y^S=1,\theta=A),\\
p(M) &\equiv \mathbb{P}(\omega^R=1\mid y^S=1,\theta=M),\\
p(U) &\equiv \mathbb{P}(\omega^R=1\mid y^S=1),
\end{align*}
where $p(U)$ is the ``unknown-type'' benchmark (the receiver observes $y^S$ but not $\theta$).%
\footnote{The experiments measure stated support. To map posteriors to observed support it is enough to assume that, in expectation, respondents are more likely to support the policy when they assign higher posterior probability to $\omega^R=1$.}

Let $\lambda\equiv \mathbb{P}(\theta=A\mid y^S=1)\in(0,1)$.\footnote{The possibility that my experimental respondents rule out the sender's cultural misalignment seems counter-intuitive. It is also inconsistent with a post-treatment question added to the Cloud Research follow-up experiment. After having exposed participants to the \textit{Pro-CPTPP} treatment condition, I asked them to guess whether the source agreed with them on abortion policy. Around \(40\%\) of the participants who made a guess replied that the source of the trade recommendation most likely disagreed with them.} By the law of total probability,
\begin{equation}\label{eq:mixture}
p(U)=\lambda\,p(A)+(1-\lambda)\,p(M).
\end{equation}
Because $p(U)$ is a strict convex combination of $p(A)$ and $p(M)$, it follows immediately that
\begin{equation}\label{eq:implication}
p(M)<p(U)\quad \Longrightarrow \quad p(A)>p(U).
\end{equation}
Thus, in the purely trust-based benchmark, negative spillovers from revealing misalignment mechanically tend to come with positive spillovers from revealing alignment. An analogous argument applies for $y^S=0$.

Empirically, misaligned bundling produces large changes in persuasion, while aligned bundling produces effects close to zero. This pattern does not rule out Bayesian reasoning in general---richer models can decouple agreement and disagreement effects---but it is difficult to reconcile with the simplest trust benchmark without adding structure beyond smooth reweighting of information.

\subsection{Identity threat and motivated distancing}\label{interpretation:identity}

A parsimonious way to organize the evidence is to distinguish between two steps that the trust benchmark compresses into one. The first is the receiver's ordinary evaluation of the economic recommendation on its merits. The second is a \emph{group-association} step: once a morally charged cue identifies the speaker as belonging to a cultural in-group or out-group, the economic recommendation can itself become encoded as a position that is ``typical'' of that group. If moral disagreement then activates identity, the receiver may distort away from positions perceived as more characteristic of the out-group. This combines the social-identity idea that threat raises identity salience \citep{TajfelWilkes,tajfel1979integrative,Oakes87,BranscombeThreat,Greenawaycruwys2018} with the belief distortions due to group identity in \citet{BGT21,GT23}. The main difference from the trust benchmark is that disagreement does not merely change how much weight the receiver places on the sender; it changes how the \emph{recommended position itself} is represented.

\paragraph{Objects.}
Let \(a\in\{0,1\}\) be an economic action available to the receiver, and let \(r\in\{0,1\}\) denote the action recommended in the message of the sender. Let \(\omega^R\in\{0,1\}\) denote the welfare-relevant state for the receiver, where action \(a=\omega^R\) is the action that is actually in the receiver's interest. After observing the economic recommendation, but \emph{before} any morally diagnostic cue is attached to the source, the receiver forms a baseline posterior
\[
\pi(a)\equiv \mathbb{P}(\omega^R=a \mid y^S=r),
\]
where \(y^S=r\) is the sender's economic recommendation. Thus \(\pi(r)\) is the receiver's posterior probability that the recommendation is correct when only the economic message is observed. Fix \(r\) and let
\[
p(U)\equiv \pi(r)
\]
denote this ``economic-only'' benchmark, and
\[
p_0 \equiv \mathbb{P}(\omega^R=r)
\]
denote the corresponding pre-message prior.

Now let \(I\) and \(O\) denote the receiver's moral in-group and out-group. For each \(j\in\{I,O\}\) and each economic action \(a\in\{0,1\}\), define
\[
q_j(a\mid \theta,c)\in(0,1), \qquad q_j(0\mid \theta,c)+q_j(1\mid \theta,c)=1,
\]
where \(q_j(a\mid \theta,c)\) is the receiver's \emph{reference belief} that a typical member of group \(j\) would endorse action \(a\), after taking into account: (i) whether the moral statement is aligned or misaligned with the receiver's views, \(\theta\in\{A,M\}\); and (ii) whether the moral and economic statements are attributed to the \emph{same} source (\(c=1\)) or to \emph{different} sources (\(c=0\)). In other words, \(q_j(a\mid \theta,c)\) measures how characteristic action \(a\) is of group \(j\) in the receiver's mental representation. It is a subjective belief about what ``people like us'' and ``people like them'' typically support. 

\paragraph{Identity-distorted updating.}
Let \(\eta(\theta)\ge 0\) denote the intensity with which moral group identity is activated by the moral cue. The conflict-driven version of the mechanism imposes
\[
\eta(A)=\eta(\varnothing)=0<\eta(M),
\]
where \(\eta(\varnothing)\) is the salience of moral group identity in the absence of any moral cue, so that only moral conflict with the source activates identity. A weaker version would only require \(\eta(\varnothing)\le\eta(A)\ll \eta(M)\), allowing small positive agreement spillovers. Given \(\pi\), \(q\), and \(\eta\), the receiver's identity-distorted posterior over economic actions is defined by
\begin{equation}\label{eq:identity_main}
\tilde{\pi}_{\theta,c}(a)
=
\pi(a) \frac{\left(\dfrac{q_I(a\mid \theta,c)}{q_O(a\mid \theta,c)}\right)^{\eta(\theta)}
}{\sum_{a'\in\{0,1\}}
\pi(a')\left(\dfrac{q_I(a'\mid \theta,c)}{q_O(a'\mid \theta,c)}\right)^{\eta(\theta)}
}.
\end{equation}
Equation \ref{eq:identity_main} adapts the identity-distortions of \citet{BGT21} to the present binary-action setting. It has a simple interpretation: the term \(\pi(a)\) captures the receiver's evaluation of the economic recommendation on its merits, while the term
\[
\left(\frac{q_I(a\mid \theta,c)}{q_O(a\mid \theta,c)}\right)^{\eta(\theta)}
\]
is an identity weight. When \(a\) is perceived as more characteristic of the in-group than of the out-group, this weight exceeds one and identity increases the appeal of \(a\). When \(a\) is perceived as more characteristic of the out-group, the weight is below one and identity reduces the appeal of \(a\). When \(\eta(\theta)=0\), the identity term disappears and the receiver relies only on the economic merits \(\pi(a)\). In this sense, the model separates a \emph{group-association margin} (captured by \(q\)) from an \emph{identity-activation margin} (captured by \(\eta\)).

For the recommended action \(r\), define
\[
p(\theta,c)\equiv \tilde{\pi}_{\theta,c}(r).
\]
Thus \(p(A,1)\) and \(p(M,1)\) are the posteriors corresponding to aligned and misaligned \emph{same-source} bundled messages, while \(p(\theta,0)\) corresponds to the separate-sources condition. These posteriors map most directly into the probability of following the recommended action. When \(r=1\), a higher posterior raises policy support; when \(r=0\), it lowers support.

Because there are only two actions, \ref{eq:identity_main} can be written in log-odds form as
\begin{equation}\label{eq:identity_logit_main}
\operatorname{logit} p(\theta,c)
=
\operatorname{logit} p(U)
+
\eta(\theta)\, d_r(\theta,c),
\end{equation}
where
\begin{equation}\label{eq:group_typing_main}
d_r(\theta,c)
\equiv
\log\!\left(
\frac{q_I(r\mid \theta,c)/q_O(r\mid \theta,c)}
     {q_I(1-r\mid \theta,c)/q_O(1-r\mid \theta,c)}
\right).
\end{equation}
The term \(d_r(\theta,c)\) is the \emph{net group-typing} of the recommended action. It is negative when the recommended action \(r\) is perceived as more characteristic of the out-group than the alternative \(1-r\), and positive when it is perceived as more characteristic of the in-group.

\paragraph{Implications.}
The model generates the paper's main empirical patterns under simple conditions. The main driving force is that when the economic recommendation and the moral statement come from the same source, they cause the economic recommendation \(r\) to become associated with that speaker's group.\footnote{If a morally misaligned speaker recommends \(r\), the receiver may begin to represent \(r\) as relatively more characteristic of the out-group than of the in-group. By contrast, when the moral and economic statements come from different sources, these group-association beliefs should move much less. This seems plausible for the kinds of issues studied here---trade and tax design---for which perceived associations with moral groups may be weak relative to high-salience moral issues.} 

First, assume that a morally misaligned speaker and a common source make the recommended action \(r\) relatively more ``out-group-typed,'' inducing the receiver to perceive the action as more typical of the outgroup:
\[
\frac{q_I(r\mid M,1)}{q_O(r\mid M,1)}
<
\frac{q_I(1-r\mid M,1)}{q_O(1-r\mid M,1)},
\]
or, equivalently, \(d_r(M,1)<0\). Since \(\eta(M)>0\), equation \ref{eq:identity_logit_main} implies
\[
p(M,1)<p(U).
\]
This is a disagreement spillover: relative to the same economic recommendation delivered without morally diagnostic content, moral disagreement reduces persuasion.

Second, the model can generate backlash relative to the no-message baseline. If the economic recommendation is persuasive on its own, so that \(p(U)>p_0\), then a misaligned same-source bundle produces backlash whenever
\[
\operatorname{logit} p(U) + \eta(M)\, d_r(M,1) < \operatorname{logit} p_0.
\]
Since \(d_r(M,1)<0\), this condition is satisfied if the negative identity term is large enough to overturn the persuasive effect of the economic recommendation.

Third, the asymmetry between disagreement and agreement is sharp in the conflict-driven version. Same-source aligned bundling may well make the recommended action more in-group-typed, so that \(d_r(A,1)>0\), but if \(\eta(A)=0\), then \(
p(A,1)=p(U)
\) and agreement does not produce a symmetric ``bonus.'' More generally, if \(\eta(A)\) is positive but small, the model predicts agreement spillovers that are weak relative to disagreement spillovers.

Fourth, under separate sources, the moral cue can still activate identity, but it no longer attaches the economic recommendation to the identity group of the speaker. The relevant restriction is therefore
\[
|d_r(M,0)|<|d_r(M,1)|.
\]
This does not sign \(d_r(M,0)\), and therefore does not sign \(p(M,0)-p(U)\). But when baseline economic group-typing is weak, so that \(d_r(M,0)\) is close to zero, the separate-sources posterior \(p(M,0)\) is correspondingly close to \(p(U)\). Thus the model predicts attenuation under source separation. This also clarifies why the moral statement alone need not move economic views: when the economic recommendation is not strongly coded as a group-typed position, there is no economic action \(r\) on which the identity distortion can operate. Hence, the mechanism is distinct from priming.

The experiments do not separately identify the reference-belief objects \(q_j(a\mid \theta,c)\) or the activation parameter \(\eta(\theta)\). But the joint pattern of the data---large negative spillovers under moral disagreement, little or no positive spillover under agreement, and sharp attenuation under source separation---is naturally captured by a model where moral conflict activates identity, and same-source bundling attaches the recommended economic action to the moral identity of the speaker.

\section{Conclusion}\label{conclusion}

Political messages rarely take a single-issue form. In contemporary U.S.\ campaigns, economic recommendations are routinely delivered alongside morally charged positions on abortion, LGBTQ rights, race, and related identity-linked issues. %This paper asks a basic question about such multi-issue communication: holding an economic argument fixed, how does its persuasiveness change when the same speaker also takes a moral stance on a salient social issue?

The paper documents a regularity. When economic recommendations are bundled with a moral stance that respondents reject, persuasion on the economic issue declines sharply, and in some cases reverses relative to a no-message baseline (backlash). When the moral stance aligns with respondents' values, however, persuasion does not reliably increase. The effect generalizes beyond abortion and trade to a second policy pair (transgender adoption and taxation) and is largely one-directional: moral disagreement spills over to economic persuasion, while economic disagreement does not comparably spill over to moral views. These findings matter for three broader reasons.

First, they identify a cross-domain channel in persuasion. Real-world persuasion often operates through \emph{bundles}: one policy stance changes how another stance is processed because the two are attributed to the same speaker. In this sense, disagreement spillovers are not merely ``another cue effect.'' They describe a specific and testable form of cross-issue interference that arises only when the moral stance is tied to the \emph{source} of the economic recommendation.

Second, they clarify how issue bundling can reshape multidimensional disagreement. A takeaway from the experiments is that bundling can shift the mapping from moral identity to economic attitudes: when misaligned receivers discount or push back against the economic recommendation, the resulting distribution of economic views becomes more stratified by moral group membership. The correlation-by-treatment evidence illustrates exactly this type of joint-distribution change within the experimental setting.

Third, they sharpen what mechanisms are plausible. A trust-based Bayesian benchmark rationalizes why moral alignment might matter—via perceived competence or perceived alignment of interests. But in its natural form, the benchmark tends to predict agreement and disagreement spillovers together. The data instead show large negative spillovers with little positive spillover. These features are naturally consistent with an identity-threat interpretation, in which moral disagreement activates conflict-driven identity salience and motivates distancing from the outgroup speaker.

The results point to several priorities for future work. What makes an issue a trigger, and what makes it a target? Moral issues such as abortion and transgender rights can act as \emph{triggers} for spillovers, while economic attitudes—especially on complex or unfamiliar policies—are comparatively \emph{targets}. A natural next step is to map more systematically the characteristics that make a policy stance likely to spill over, including identity relevance, perceived importance, or social desirability. Answering this would move toward a general taxonomy of spillover triggers and targets.

The identity-threat channel has predictions that can be measured and manipulated more directly: perceived outgroup hostility, identity salience, anger/affect, and counter-arguing after exposure. Relatedly, one could vary the connotation of the moral stance (e.g., hostile attack vs.\ principled disagreement) while holding the policy position fixed, to test whether spillovers respond to threat intensity.

More broadly, disagreement spillovers suggest that moral conflict can constrain economic persuasion: economic arguments may fail---or even backfire---when bundled with moral stances that activate outgroup antagonism. Understanding when and how these spillovers can be deactivated is therefore important both for positive models of polarization and normative questions about communication in divided societies.

\newpage \bibliographystyle{apalike} \bibliography{biblio} \newpage \appendix \setcounter{table}{0} \renewcommand{\thetable}{A\arabic{table}} \vspace{1cm}

@article{BGT21,
    author = {Bonomi, Giampaolo and Gennaioli, Nicola and Tabellini, Guido},
    title = "{Identity, Beliefs, and Political Conflict}",
    journal = {The Quarterly Journal of Economics},
    volume = {136},
    number = {4},
    pages = {2371-2411},
    year = {2021},
    month = {09},
    issn = {0033-5533},
    doi = {10.1093/qje/qjab034},
    url = {https://doi.org/10.1093/qje/qjab034},
    eprint = {https://academic.oup.com/qje/article-pdf/136/4/2371/40566470/qjab034.pdf},
}

@article{GT23,
    author = {Gennaioli, Nicola and Tabellini, Guido},
    title = "{Identity Politics}",
    journal = {Working Paper},
   year= {2023}
}

@article{Fukuyama2018,
 ISSN = {00157120},
 URL = {http://www.jstor.org/stable/44823914},
 author = {Francis Fukuyama},
 journal = {Foreign Affairs},
 number = {5},
 pages = {90--114},
 publisher = {Council on Foreign Relations},
 title = {Against Identity Politics: The New Tribalism and the Crisis of Democracy},
 urldate = {2024-04-03},
 volume = {97},
 year = {2018}
}

@book{Norris_Inglehart_2019, place={Cambridge}, title={Cultural Backlash: Trump, Brexit, and Authoritarian Populism}, publisher={Cambridge University Press}, author={Norris, Pippa and Inglehart, Ronald}, year={2019}}

@article{Gentzkow2016,
  title={Polarization in 2016},
  author={Matthew Gentzkow},
  year={2016},
journal = {Working Paper},
  url={https://api.semanticscholar.org/CorpusID:157386121}
}

@article{Cohen2003,
  title={Party over policy: The dominating impact of group influence on political beliefs.},
  author={Cohen, Geoffrey L},
  journal={Journal of personality and social psychology},
  volume={85},
  number={5},
  pages={808},
  year={2003},
  publisher={American Psychological Association}}

@article{Feddersen2022,
title = {Public opinion backlash in response to party messages: A case for party press releases},
journal = {Electoral Studies},
volume = {80},
pages = {102532},
year = {2022},
issn = {0261-3794},
doi = {https://doi.org/10.1016/j.electstud.2022.102532},
url = {https://www.sciencedirect.com/science/article/pii/S0261379422000889},
author = {Alexandra Feddersen and James F. Adams}
}

@article{Druckman,
   author = "Druckman, James N.",
   title = "A Framework for the Study of Persuasion", 
   journal= "Annual Review of Political Science",
   year = "2022",
   volume = "25",
   number = "Volume 25, 2022",
   pages = "65-88",
   doi = "https://doi.org/10.1146/annurev-polisci-051120-110428",
   url = "https://www.annualreviews.org/content/journals/10.1146/annurev-polisci-051120-110428",
   publisher = "Annual Reviews",
   issn = "1545-1577",
   type = "Journal Article",
  }

@article{Enke2022,
    author = {Enke, Benjamin and Rodríguez-Padilla, Ricardo and Zimmermann, Florian},
    title = "{Moral Universalism and the Structure of Ideology}",
    journal = {The Review of Economic Studies},
    volume = {90},
    number = {4},
    pages = {1934-1962},
    year = {2022},
    month = {09},
    issn = {0034-6527},
    doi = {10.1093/restud/rdac066},
    url = {https://doi.org/10.1093/restud/rdac066},
    eprint = {https://academic.oup.com/restud/article-pdf/90/4/1934/50827247/rdac066.pdf},
}

@article{Enke2024,
   author = "Enke, Benjamin",
   title = "Moral Boundaries", 
   journal= "Annual Review of Economics",
   year = "2024",
   volume = "16",
   number = "Volume 16, 2024",
   pages = "133-157",
   doi = "https://doi.org/10.1146/annurev-economics-091223-093730",
   url = "https://www.annualreviews.org/content/journals/10.1146/annurev-economics-091223-093730",
   publisher = "Annual Reviews",
   issn = "1941-1391",
   type = "Journal Article",
  }

@techreport{Enke2022values,
 title = "Values as Luxury Goods and Political Polarization",
 author = "Enke, Benjamin and Polborn, Mattias and Wu, Alex",
 institution = "National Bureau of Economic Research",
 type = "Working Paper",
 series = "Working Paper Series",
 number = "30001",
 year = "2022",
 month = "April",
 doi = {10.3386/w30001},
 URL = "http://www.nber.org/papers/w30001",
}

@article{Enke2020,
author = {Enke, Benjamin},
title = {Moral Values and Voting},
journal = {Journal of Political Economy},
volume = {128},
number = {10},
pages = {3679-3729},
year = {2020},
doi = {10.1086/708857},

URL = { 
    
        https://doi.org/10.1086/708857
    
    

},
eprint = { 
    
        https://doi.org/10.1086/708857
}
}

@article{Parker13,
    author = {Parker, M.T. and Janoff-Bulman, R.},
    title = {Lessons from Morality-Based Social Identity: The Power of Outgroup ``Hate," Not Just Ingroup ``Love"},
    journal = {Social Justice Research},
    volume = {26},
    pages = {81-96},
    year = {2013},
}

@article{Afrouzi2024,
author = {Afrouzi, Hassan and Arteaga, Carolina and Weisburst, Emily},
title = {Is It the Message or the Messenger? Examining Movement in Immigration Beliefs},
journal = {Journal of Political Economy Microeconomics},
volume = {2},
number = {2},
pages = {244-297},
year = {2024},
doi = {10.1086/728365},

URL = { 
    
        https://doi.org/10.1086/728365
    
    

},
eprint = { 
    
        https://doi.org/10.1086/728365
    
    

}
}

@article{BarberPope19, title={Does Party Trump Ideology? Disentangling Party and Ideology in America}, volume={113}, DOI={10.1017/S0003055418000795}, number={1}, journal={American Political Science Review}, author={Barber, Michael and Pope, Jeremy C.}, year={2019}, pages={38–54}}

@article{ReyBiel, title={Ideological Alignment and Evidence-Based Policy Adoption}, journal={Working Paper}, author={Jansen, Marcel and Martinez, Angel and Ozkan, Berkay and Rey-Biel, Pedro and Roldan-Mones, Antonio}, year={2024}}

@article{Gethin,
    author = {Gethin, Amory and Martínez-Toledano, Clara and Piketty, Thomas},
    title = "{Brahmin Left Versus Merchant Right: Changing Political Cleavages in 21 Western Democracies, 1948–2020*}",
    journal = {The Quarterly Journal of Economics},
    volume = {137},
    number = {1},
    pages = {1-48},
    year = {2021},
    month = {10},
    issn = {0033-5533},
    doi = {10.1093/qje/qjab036},
    url = {https://doi.org/10.1093/qje/qjab036},
    eprint = {https://academic.oup.com/qje/article-pdf/137/1/1/42193968/qjab036.pdf},
}

@article{Desmet, title={Endogenous Partitions!}, journal={Working Paper}, author={Desmet, Klaus and Ortuño-Ortín, Ignacio and Romain Wacziarg}, year={2024}}

@article{BesleyPersson, author = {Besley, Timothy and Persson, Torsten}, title = {The Rise of Identity Politics: Policy, Political Organization, and Nationalist Dynamics}, journal = {Working Paper}, year = {2021}}

@TECHREPORT{Romer2007,
title = {Racism, Xenophobia, and Distribution: Multi-issue Politics in Advanced Democracies},
author = {Roemer, John and Woojin, Lee and Van der Straeten, Karine},
year = {2007},
institution = {HAL},
type = {Post-Print},
url = {https://EconPapers.repec.org/RePEc:hal:journl:halshs-00754747}
}

@article{Gentzkow2024,
  title={Ideological Bias and Trust in Information Sources},
  author={Matthew Gentzkow and Michael B. T. Wong and Allen Zhang},
  year={2024},
 journal = {AEJ: Microeconomics (forthcoming)}
}

@article{tajfel1979integrative,
  author = {Tajfel, Henri and Turner, John C},
  journal = {The Social Psychology of Intergroup Relations},
  number = 47,
  pages = 74,
  title = {An Integrative Theory of Intergroup Conflict},
  url = {http://scholar.google.de/scholar.bib?q=info:_JZ4OXw0-7oJ:scholar.google.com/&output=citation&hl=en&ct=citation&cd=0},
  volume = 33,
  year = 1979
}

@article{TajfelWilkes,
author = {Tajfel, Henry and Wilkes, A. L.},
title = {Classification and Quantitative Judgement},
journal = {British Journal of Psychology},
volume = {54},
number = {2},
pages = {101-114},
doi = {https://doi.org/10.1111/j.2044-8295.1963.tb00865.x},
url = {https://bpspsychub.onlinelibrary.wiley.com/doi/abs/10.1111/j.2044-8295.1963.tb00865.x},
eprint = {https://bpspsychub.onlinelibrary.wiley.com/doi/pdf/10.1111/j.2044-8295.1963.tb00865.x},
year = {1963}
}

@inproceedings{Branscombethreat,
  title={The context and content of social identity threat},
  booktitle={Social identity: Context, commitment, content},
  author={Nyla R. Branscombe and Naomi Ellemers and Russell Spears and E. J. Doosje},
  year={1999},
  url={https://api.semanticscholar.org/CorpusID:70942483}
}

@misc{Greenawaycruwys2018, title={The source model of group threat: Responding to internal and external threats}, url={osf.io/preprints/psyarxiv/k2t56}, DOI={10.31234/osf.io/k2t56}, publisher={PsyArXiv}, author={Greenaway, Katharine H and Cruwys, Tegan}, year={2018}, month={May}}

@incollection{Oakes87,
  author      = "Oakes, Penelope J.",
  title       = "The Salience of Social Categories",
  booktitle   = "Rediscovering the Social Group: A
Self-Categorization Theory",
  publisher   = "Oxford: Blackwell",
editor = {Turner, John C. and  Hogg, Michael A. and Oakes, Penelope J. and J. Reicher, Stephen
 and Wetherell, Margaret J.},
  year        = 1987
}

@article{Kunda1990TheCF,
  title={The case for motivated reasoning.},
  author={Ziva Kunda},
  journal={Psychological bulletin},
  year={1990},
  volume={108 3},
  pages={
          480-98
        },
  url={https://api.semanticscholar.org/CorpusID:9703661}
}

@misc{bonomi2024aearctr12729,
  author       = {Bonomi, Giampaolo},
  title        = {Information Bundling and Polarizing Persuasion},
  year         = {2024},
  howpublished = {AEA RCT Registry},
  institution  = {American Economic Association},
  note         = {RCT ID: AEARCTR-0012729.},
  url          = {https://www.socialscienceregistry.org/trials/12729},
  urldate      = {2026-03-09}
}

% ==========================================================
% APPENDIX
% ==========================================================
\appendix
\setcounter{table}{0}
\renewcommand{\thetable}{A\arabic{table}}
\setcounter{figure}{0}
\renewcommand{\thefigure}{A\arabic{figure}}

% ==========================================================
% APPENDIX A: EMPIRICAL APPENDIX (implementation + extra tables)
% ==========================================================
\section{Empirical Appendix}\label{empapp}

\subsection{Survey structure, instruments, and randomization}\label{app:surveystructure}

\paragraph{Overview of modules (abortion--trade experiment).}
The abortion--trade survey consists of five blocks:
(i) brief neutral background descriptions of the two policy topics;
(ii) pre-treatment measurement of abortion views and pre-treatment covariates;
(iii) randomized exposure to one of seven message arms;
(iv) post-treatment measurement of CPTPP support and related outcomes;
(v) remaining demographics and debrief.

\paragraph{Policy background descriptions.}
Before any outcome measurement, respondents read short neutral descriptions of (a) abortion laws and (b) the CPTPP. These descriptions are informational and do not include partisan labels or endorsements.

\paragraph{Pre-treatment measures.}
Respondents then report baseline abortion views \emph{before} treatment.
The main baseline question is:
\begin{quote}
``Do you support or oppose laws imposing strict restrictions on abortion / making abortion illegal?''
\end{quote}
Responses are collected on a four-point scale (strongly oppose / somewhat oppose / somewhat support / strongly support).
For analysis, respondents are classified as \textit{pro-choice} if they answer oppose (either category) and \textit{pro-life} if they answer support (either category).

Additional pre-treatment covariates include demographics (e.g. age, gender, race, education, employment) and party identity (five-point scale), as well as self-reported importance/knowledge items for abortion and trade.

\paragraph{Treatment assignment.}
Respondents are randomized at the respondent level (Qualtrics randomizer) into seven arms with approximately equal allocation:
(i) no-message control;
(ii) pro-CPTPP message only;
(iii) anti-CPTPP message only;
(iv)--(vii) bundled arms pairing each trade recommendation with either a pro-life or pro-choice abortion stance.
In bundled arms, the order of passages (abortion first vs.\ trade first) is randomized to mitigate order/recency concerns.

\paragraph{Post-treatment measures.}
After message exposure, respondents answer:
\begin{quote}
``Do you support or oppose U.S.\ participation in the CPTPP?''
\end{quote}
on the same four-point scale. The main binary outcome $Y_i$ codes support (somewhat or strongly support) as 1 and oppose (somewhat or strongly oppose) as 0. Ordered-response robustness is reported in Table~\ref{ordered}.

\paragraph{Attention checks and exclusions.}
The main analysis excludes (i) respondents flagged by Qualtrics ex-post fraud/bot detection and (ii) respondents failing two attention checks designed to verify that they read the stimulus text. These restrictions remove approximately 5\% of completed responses and have almost no effect on the estimated coefficient of interest. Balance checks after exclusions are reported in Table~\ref{balance}.

\paragraph{Follow-up design for Table~\ref{nopty}.}
After the main wave, a subset of participants who had not previously received a trade message were re-contacted for a short follow-up. The follow-up is a within-respondent pre/post design: respondents report CPTPP support, then read one of three message conditions, then report CPTPP support again. The three conditions are (i) pro-CPTPP only; (ii) pro-life + pro-CPTPP bundled, with the speaker described as non-partisan; (iii) the same two passages presented as coming from two unrelated, non-partisan speakers (Separate Messages). Table~\ref{nopty} reports treatment effects on the within-respondent change in CPTPP support, with standard errors clustered by respondent.

\paragraph{Preregistration and fielding dates.}
The main CloudResearch experiment was preregistered in the AEA RCT Registry (AEARCTR-0012729) before data collection began. Data were collected between July and August 2024. The Prolific experiment implementing the second policy pair was fielded in August 2024.

\newpage
\subsection{Treatment texts}\label{app:messages}

For convenience, this section lists the exact message text used in each experiment.

\subsubsection{Abortion--trade experiment}\label{app:messages_trade}

\paragraph{No message.} Respondents do not read any political message.

\paragraph{Pro-CPTPP message.}
\textit{``Regarding trade policy, joining the CPTPP would undeniably boost our economy, create new jobs, and lower consumer prices. The US should join the CPTPP, and do it as soon as possible.''}

\paragraph{Anti-CPTPP message.}
\textit{``Regarding the CPTPP, the trade deal is by and large against the interest of our country. It would destroy US jobs and increase economic insecurity. The US should stay out of the CPTPP.''}

\paragraph{Pro-life, pro-CPTPP message.}
\textit{``Abortion cannot be allowed as a free choice, it must be either illegal or strongly discouraged by law. Exceptional situations might need special attention, but the life of the unborn is a gift of God and needs to be protected. [...] Regarding trade policy, joining the CPTPP would undeniably boost our economy, create new jobs, and lower consumer prices. The US should join the CPTPP, and do it as soon as possible.''}

\paragraph{Pro-choice, pro-CPTPP message.}
\textit{``Abortion should always be allowed, with no restrictions, as a free and personal choice. The right to abortion is implicit in the concept of liberty, and any attempt to dictate to women what to do with their bodies is unacceptable. [...] Regarding trade policy, joining the CPTPP would undeniably boost our economy, create new jobs, and lower consumer prices. The US should join the CPTPP, and do it as soon as possible.''}

\paragraph{Pro-life, anti-CPTPP message.}
\textit{``Abortion cannot be allowed as a free choice, it must be either illegal or strongly discouraged by law. Exceptional situations might need special attention, but the life of the unborn is a gift of God and needs to be protected. [...] Regarding the CPTPP, the trade deal is by and large against the interest of our country. It would destroy US jobs and increase economic insecurity. The US should stay out of the CPTPP.''}

\paragraph{Pro-choice, anti-CPTPP message.}
\textit{``Abortion should always be allowed, with no restrictions, as a free and personal choice. The right to abortion is implicit in the concept of liberty, and any attempt to dictate to women what to do with their bodies is unacceptable. [...] Regarding the CPTPP, the trade deal is by and large against the interest of our country. It would destroy US jobs and increase economic insecurity. The US should stay out of the CPTPP.''}

\subsubsection{Transgender adoption--taxation experiment}\label{app:messages_tax}

\paragraph{Pro-tax-reform message.}
\textit{``Regarding taxation, a brave and sensible policy would be to cut income taxes for everyone, dramatically, and compensate by raising the VAT. This would be fairer, as it gives people more control over how much they will pay, and would also be sustainable.''}

\paragraph{Pro-adoption-ban, pro-tax-reform message.}
\textit{``It takes a biological male and a biological female to conceive a child, and this makes the traditional family naturally predisposed to raise children. Adoption by members of the LGBTQ community is against the rules of nature, and, in the case of transgender individuals, it should be banned. [...] Regarding taxation, a brave and sensible policy would be to cut income taxes for everyone, dramatically, and compensate by raising the VAT. This would be more fair, as it gives people more control over how much they will pay and would also be sustainable.''}

\newpage

\subsection{Experimental Analysis: Supplementary Tables}\label{apptables} 
% Table generated by Excel2LaTeX from sheet 'Balance Tables'
\begin{table}[h!]
\begin{adjustbox}{width = .8\textwidth}
  \caption{Balance Tests}\label{balance}
  \centering
    \begin{tabular}{p{10.75em}llllllll}
    \toprule
    \multicolumn{1}{r}{} &       &       &       &       &       &       &       &  \\
    \multicolumn{1}{r}{} &       &       & \multicolumn{2}{p{10em}}{Pro-CPTPP Message} &       &       & \multicolumn{2}{p{10em}}{Anti-CPTPP Message} \\
\cmidrule{4-5}\cmidrule{8-9}    \multicolumn{1}{r}{} &       &       & \multicolumn{1}{c}{mean} & \multicolumn{1}{c}{ p-value} &       &       & \multicolumn{1}{c}{mean} & \multicolumn{1}{c}{p-value} \\
    \midrule
    \multicolumn{1}{r}{} &       &       &       &       &       &       &       &  \\
    \multicolumn{1}{l}{Pro life} &       &       & \multicolumn{1}{c}{0.276} & \multicolumn{1}{c}{0.979} &       &       & \multicolumn{1}{c}{0.249} & \multicolumn{1}{c}{0.503} \\
    \multicolumn{1}{l}{5-Point party scale} &       &       & \multicolumn{1}{c}{3.329} & \multicolumn{1}{c}{0.626} &       &       & \multicolumn{1}{c}{3.293} & \multicolumn{1}{c}{0.297} \\
    \multicolumn{1}{l}{Male} &       &       & \multicolumn{1}{c}{0.438} & \multicolumn{1}{c}{0.470} &       &       & \multicolumn{1}{c}{0.456} & \multicolumn{1}{c}{0.392} \\
    \multicolumn{1}{l}{Upper-middle/upper class} &       &       & \multicolumn{1}{c}{0.309} & \multicolumn{1}{c}{0.425} &       &       & \multicolumn{1}{c}{0.278} & \multicolumn{1}{c}{0.497} \\
    \multicolumn{1}{l}{Lower class} &       &       & \multicolumn{1}{c}{0.162} & \multicolumn{1}{c}{0.507} &       &       & \multicolumn{1}{c}{0.173} & \multicolumn{1}{c}{0.942} \\
    \multicolumn{1}{l}{Lower-middle class} &       &       & \multicolumn{1}{c}{0.529} & \multicolumn{1}{c}{0.796} &       &       & \multicolumn{1}{c}{0.550} & \multicolumn{1}{c}{0.350} \\
    \multicolumn{1}{l}{High school or less} &       &       & \multicolumn{1}{c}{0.103} & \multicolumn{1}{c}{0.830} &       &       & \multicolumn{1}{c}{0.099} & \multicolumn{1}{c}{0.246} \\
    \multicolumn{1}{l}{Some college} &       &       & \multicolumn{1}{c}{0.219} & \multicolumn{1}{c}{0.376} &       &       & \multicolumn{1}{c}{0.218} & \multicolumn{1}{c}{0.120} \\
    \multicolumn{1}{l}{College degree} &       &       & \multicolumn{1}{c}{0.463} & \multicolumn{1}{c}{0.529} &       &       & \multicolumn{1}{c}{0.472} & \multicolumn{1}{c}{0.496} \\
    \multicolumn{1}{l}{Postgraduate degree} &       &       & \multicolumn{1}{c}{0.213} & \multicolumn{1}{c}{0.525} &       &       & \multicolumn{1}{c}{0.208} & \multicolumn{1}{c}{0.527} \\
    \multicolumn{1}{l}{Unemployed} &       &       & \multicolumn{1}{c}{0.097} & \multicolumn{1}{c}{0.323} &       &       & \multicolumn{1}{c}{0.104} & \multicolumn{1}{c}{0.311} \\
    \multicolumn{1}{l}{Employed} &       &       & \multicolumn{1}{c}{0.780} & \multicolumn{1}{c}{0.274} &       &       & \multicolumn{1}{c}{0.777} & \multicolumn{1}{c}{0.658} \\
    \multicolumn{1}{l}{Other employment status} &       &       & \multicolumn{1}{c}{0.123} & \multicolumn{1}{c}{0.805} &       &       & \multicolumn{1}{c}{0.119} & \multicolumn{1}{c}{0.989} \\
    \multicolumn{1}{l}{Religion important} &       &       & \multicolumn{1}{c}{0.353} & \multicolumn{1}{c}{0.685} &       &       & \multicolumn{1}{c}{0.338} & \multicolumn{1}{c}{0.361} \\
    \multicolumn{1}{l}{Atheist/Agnostic} &       &       & \multicolumn{1}{c}{0.352} & \multicolumn{1}{c}{0.964} &       &       & \multicolumn{1}{c}{0.365} & \multicolumn{1}{c}{0.977} \\
    \multicolumn{1}{l}{Christian, evangelical} &       &       & \multicolumn{1}{c}{0.147} & \multicolumn{1}{c}{0.503} &       &       & \multicolumn{1}{c}{0.140} & \multicolumn{1}{c}{0.481} \\
    \multicolumn{1}{l}{Jewish} &       &       & \multicolumn{1}{c}{0.023} & \multicolumn{1}{c}{0.187} &       &       & \multicolumn{1}{c}{0.017} & \multicolumn{1}{c}{0.303} \\
    \multicolumn{1}{l}{Other religion} &       &       & \multicolumn{1}{c}{0.114} & \multicolumn{1}{c}{0.603} &       &       & \multicolumn{1}{c}{0.129} & \multicolumn{1}{c}{0.118} \\
    \multicolumn{1}{l}{Christian, other} &       &       & \multicolumn{1}{c}{0.354} & \multicolumn{1}{c}{0.945} &       &       & \multicolumn{1}{c}{0.346} & \multicolumn{1}{c}{0.738} \\
    \multicolumn{1}{l}{Black} &       &       & \multicolumn{1}{c}{0.091} & \multicolumn{1}{c}{0.201} &       &       & \multicolumn{1}{c}{0.094} & \multicolumn{1}{c}{0.166} \\
    \multicolumn{1}{l}{Hispanic} &       &       & \multicolumn{1}{c}{0.060} & \multicolumn{1}{c}{0.892} &       &       & \multicolumn{1}{c}{0.059} & \multicolumn{1}{c}{0.262} \\
    \multicolumn{1}{l}{White} &       &       & \multicolumn{1}{c}{0.760} & \multicolumn{1}{c}{0.547} &       &       & \multicolumn{1}{c}{0.765} & \multicolumn{1}{c}{0.921} \\
    \multicolumn{1}{l}{Age: 18-35} &       &       & \multicolumn{1}{c}{0.295} & \multicolumn{1}{c}{0.132} &       &       & \multicolumn{1}{c}{0.301} & \multicolumn{1}{c}{0.379} \\
    \multicolumn{1}{l}{Age: 36-50} &       &       & \multicolumn{1}{c}{0.407} & \multicolumn{1}{c}{0.066} &       &       & \multicolumn{1}{c}{0.407} & \multicolumn{1}{c}{0.711} \\
    \multicolumn{1}{l}{Age: 51-65} &       &       & \multicolumn{1}{c}{0.212} & \multicolumn{1}{c}{0.518} &       &       & \multicolumn{1}{c}{0.221} & \multicolumn{1}{c}{0.276} \\
    \multicolumn{1}{l}{Age: 66+} &       &       & \multicolumn{1}{c}{0.085} & \multicolumn{1}{c}{0.612} &       &       & \multicolumn{1}{c}{0.070} & \multicolumn{1}{c}{0.637} \\
    \multicolumn{1}{l}{Abortion knowledge} &       &       & \multicolumn{1}{c}{7.721} & \multicolumn{1}{c}{0.824} &       &       & \multicolumn{1}{c}{7.629} & \multicolumn{1}{c}{0.602} \\
    \multicolumn{1}{l}{Trade knowledge} &       &       & \multicolumn{1}{c}{4.545} & \multicolumn{1}{c}{0.964} &       &       & \multicolumn{1}{c}{4.346} & \multicolumn{1}{c}{0.755} \\
    \multicolumn{1}{l}{Abortion Importance} &       &       & \multicolumn{1}{c}{7.452} & \multicolumn{1}{c}{0.774} &       &       & \multicolumn{1}{c}{7.444} & \multicolumn{1}{c}{0.225} \\
    \multicolumn{1}{l}{Trade Importance} &       &       & \multicolumn{1}{c}{5.414} & \multicolumn{1}{c}{0.690} &       &       & \multicolumn{1}{c}{5.400} & \multicolumn{1}{c}{0.630} \\
    \multicolumn{1}{r}{} &       &       &       &       &       &       &       &  \\
    \midrule
    \midrule
    \multicolumn{9}{p{38em}}{\small Notes: For each type of trade message and demographic characteristic, the table reports (i) the mean of the demographic characteristic in the sample is comprised of aligned, misaligned, trade-only, and no-message respondents; (ii) the p-value of the F-test of joint significance of a regression of the demographic characteristic on group indicators for aligned, misaligned, trade-only, and no-message respondents (the omitted category). Aligned respondents received the corresponding trade recommendation bundled together with an abortion message aligned with their views. Misaligned respondents received the corresponding trade recommendation together with an abortion message in contrast with their views. Trade-only respondents read the corresponding trade recommendation without any additional abortion content. No-message respondents were part of the treatment arm that was not shown any political message. All demographic variables are indicators for whether the respondent exhibits the corresponding characteristic, except for party scale, which is a 5-point partisanship scale (higher values denoting Democratic identity), abortion/Trade knowledge, and abortion/trade importance, which are on 11-point scales.} \\
    \end{tabular}%
\end{adjustbox}
\end{table}%
\newpage 
\begin{table}[h!]
\begin{adjustbox}{width = \textwidth}
  \centering
  \caption{Aggregate Effect (Backlash) Breakdown}\label{AggrBreak}
    \begin{tabular}{p{13.585em}llllll}
    \toprule
    \multicolumn{1}{r}{} &       &       &       &       &       &  \\
    \multicolumn{1}{r}{} &       & \multicolumn{5}{c}{Differences in Share of} \\
    \multicolumn{1}{r}{} &       & \multicolumn{5}{c}{ Respondents Supporting the CPTPP} \\
    \multicolumn{1}{r}{} &       & \multicolumn{5}{c}{(Relative to No Message)} \\
    \multicolumn{1}{r}{} &       &       &       &       &       &  \\
    \multicolumn{1}{r}{} &       & \multicolumn{2}{c}{Pro Life Respondents} &       & \multicolumn{2}{c}{Pro-Choice Respondents} \\
    \midrule
    \multicolumn{1}{r}{} &       &       &       &       &       &  \\
    \multicolumn{1}{l}{Pro-Life, Pro-CPTPP Message} &       & \multicolumn{1}{c}{0.0571} & \multicolumn{1}{c}{0.0768} &       & \multicolumn{1}{c}{-0.112***} & \multicolumn{1}{c}{-0.109***} \\
    \multicolumn{1}{r}{} &       & \multicolumn{1}{c}{(0.0517)} & \multicolumn{1}{c}{(0.0528)} &       & \multicolumn{1}{c}{(0.0303)} & \multicolumn{1}{c}{(0.0308)} \\
    \multicolumn{1}{r}{} &       & \multicolumn{1}{c}{[262]} & \multicolumn{1}{c}{[262]} &       & \multicolumn{1}{c}{[680]} & \multicolumn{1}{c}{[678]} \\
    \multicolumn{1}{r}{} &       &       &       &       &       &  \\
    \multicolumn{1}{l}{Pro-Life, Anti-CPTPP Message} &       & \multicolumn{1}{c}{-0.529***} & \multicolumn{1}{c}{-0.490***} &       & \multicolumn{1}{c}{-0.231***} & \multicolumn{1}{c}{-0.225***} \\
    \multicolumn{1}{r}{} &       & \multicolumn{1}{c}{(0.0663)} & \multicolumn{1}{c}{(0.0720)} &       & \multicolumn{1}{c}{(0.0388)} & \multicolumn{1}{c}{(0.0399)} \\
    \multicolumn{1}{r}{} &       & \multicolumn{1}{c}{[177]} & \multicolumn{1}{c}{[177]} &       & \multicolumn{1}{c}{[456]} & \multicolumn{1}{c}{[455]} \\
    \multicolumn{1}{r}{} &       &       &       &       &       &  \\
    \multicolumn{1}{l}{Pro-Choice, Pro-CPTPP Message} &       & \multicolumn{1}{c}{-0.141**} & \multicolumn{1}{c}{-0.164**} &       & \multicolumn{1}{c}{0.0465} & \multicolumn{1}{c}{0.0448} \\
    \multicolumn{1}{r}{} &       & \multicolumn{1}{c}{(0.0683)} & \multicolumn{1}{c}{(0.0712)} &       & \multicolumn{1}{c}{(0.0315)} & \multicolumn{1}{c}{(0.0312)} \\
    \multicolumn{1}{r}{} &       & \multicolumn{1}{c}{[179]} & \multicolumn{1}{c}{[178]} &       & \multicolumn{1}{c}{[454]} & \multicolumn{1}{c}{[452]} \\
    \multicolumn{1}{r}{} &       &       &       &       &       &  \\
    \multicolumn{1}{l}{Pro-Choice, Anti-CPTPP Message} &       & \multicolumn{1}{c}{-0.439***} & \multicolumn{1}{c}{-0.410***} &       & \multicolumn{1}{c}{-0.471***} & \multicolumn{1}{c}{-0.463***} \\
    \multicolumn{1}{r}{} &       & \multicolumn{1}{c}{(0.0724)} & \multicolumn{1}{c}{(0.0765)} &       & \multicolumn{1}{c}{(0.0380)} & \multicolumn{1}{c}{(0.0392)} \\
    \multicolumn{1}{r}{} &       & \multicolumn{1}{c}{[153]} & \multicolumn{1}{c}{[153]} &       & \multicolumn{1}{c}{[495]} & \multicolumn{1}{c}{[494]} \\
    \multicolumn{1}{r}{} &       &       &       &       &       &  \\
    \multicolumn{1}{l}{Demographic Controls} &       & \multicolumn{1}{c}{N} & \multicolumn{1}{c}{Y} &       & \multicolumn{1}{c}{N} & \multicolumn{1}{c}{Y} \\
    \multicolumn{1}{r}{} &       &       &       &       &       &  \\
    \midrule
    \midrule
    \multicolumn{7}{p{38.67em}}{Notes: The dependent variable is an indicator variable equal to 1 if the respondent supports the US membership in the CPTPP. Coefficients are differences relative to participants in the treatment arm where no message is sent. Demographic controls include three age categories, and indicators for whether the respondent identifies as a man, is white, is Christian,  is a Republican or a Democrat,  has a college degree, identifies as upper-class or upper-middle class, and is unemployed.  Robust standard errors are in parentheses. Sample size in square brackets.  *** \(p<0.01\), ** \(p<0.05\), *\(p<0.1\)} \\
    \end{tabular}%
\end{adjustbox}
\end{table}%
\newpage 

\begin{table}[htbp]
  \centering
  \caption{Spillovers and Backlash from Abortion to Trade (Time Polynomial Spline)}\label{spline}
\begin{adjustbox}{width=\textwidth}
    \begin{tabular}{p{13em}llllll}
    \toprule
    \multicolumn{1}{r}{} &       &       &       &       &       &  \\
    \multicolumn{1}{r}{} &       & \multicolumn{5}{c}{Differences in Share of} \\
    \multicolumn{1}{r}{} &       & \multicolumn{5}{c}{ Respondents Supporting the CPTPP} \\
    \multicolumn{1}{r}{} &       & \multicolumn{2}{c}{} &       & \multicolumn{2}{c}{} \\
    \midrule
    \multicolumn{1}{r}{} &       &       &       &       &       &  \\
    \multicolumn{1}{r}{} &       & \multicolumn{5}{c}{Panel A. Pro-CPTPP Message } \\
    \multicolumn{1}{r}{} &       &       &       &       &       &  \\
    \multicolumn{1}{r}{} &       & \multicolumn{2}{c}{Relative to} &       & \multicolumn{2}{c}{Relative to} \\
    \multicolumn{1}{r}{} &       & \multicolumn{2}{c}{Pro-CPTPP Only} &       & \multicolumn{2}{c}{No Message} \\
\cmidrule{3-7}    \multicolumn{1}{l}{Pro-CPTPP, Aligned Abortion (\(\tau_A^{\text{Pro}}\))} &       & \multicolumn{1}{c}{0.0233 } & \multicolumn{1}{c}{0.0233 } &       & \multicolumn{1}{c}{0.0445 } & \multicolumn{1}{c}{0.0473* } \\
    \multicolumn{1}{r}{} &       & \multicolumn{1}{c}{(0.0225)} & \multicolumn{1}{c}{(0.0220)} &       & \multicolumn{1}{c}{(0.0288)} & \multicolumn{1}{c}{(0.0287)} \\
    \multicolumn{1}{l}{Pro-CPTPP, Misaligned Abortion (\(\tau_M^{\text{Pro}}\))} &       & \multicolumn{1}{c}{-0.127***} & \multicolumn{1}{c}{-0.127***} &       & \multicolumn{1}{c}{-0.108***} & \multicolumn{1}{c}{-0.107***} \\
    \multicolumn{1}{r}{} &       & \multicolumn{1}{c}{(0.0271)} & \multicolumn{1}{c}{(0.0271)} &       & \multicolumn{1}{c}{(0.0325)} & \multicolumn{1}{c}{(0.0328)} \\
    \multicolumn{1}{r}{} &       &       &       &       &       &  \\
    \multicolumn{1}{l}{Observations} &       & \multicolumn{1}{c}{1254 } & \multicolumn{1}{c}{1251 } &       & \multicolumn{1}{c}{1050 } & \multicolumn{1}{c}{1047 } \\
    \multicolumn{1}{r}{} &       &       &       &       &       &  \\
    \midrule
    \multicolumn{1}{r}{} &       &       &       &       &       &  \\
    \multicolumn{1}{r}{} &       & \multicolumn{5}{c}{Panel B. Anti-CPTPP Message } \\
    \multicolumn{1}{r}{} &       &       &       &       &       &  \\
    \multicolumn{1}{r}{} &       & \multicolumn{2}{c}{Relative to} &       & \multicolumn{2}{c}{Relative to} \\
    \multicolumn{1}{r}{} &       & \multicolumn{2}{c}{Anti-CPTPP Only} &       & \multicolumn{2}{c}{No Message} \\
\cmidrule{3-7}    \multicolumn{1}{l}{Anti-CPTPP, Aligned Abortion (\(\tau_A^{\text{Anti}}\))} &       & \multicolumn{1}{c}{0.0261} & \multicolumn{1}{c}{0.0297} &       & \multicolumn{1}{c}{-0.462***} & \multicolumn{1}{c}{-0.449***} \\
    \multicolumn{1}{r}{} &       & \multicolumn{1}{c}{(0.0385)} & \multicolumn{1}{c}{(0.0376)} &       & \multicolumn{1}{c}{(0.0386)} & \multicolumn{1}{c}{(0.0386)} \\
    \multicolumn{1}{l}{Anti-CPTPP, Misaligned Abortion (\(\tau_M^{\text{Anti}}\))} &       & \multicolumn{1}{c}{0.213***} & \multicolumn{1}{c}{0.209***} &       & \multicolumn{1}{c}{-0.269***} & \multicolumn{1}{c}{-0.267***} \\
    \multicolumn{1}{r}{} &       & \multicolumn{1}{c}{(0.0347)} & \multicolumn{1}{c}{(0.0340)} &       & \multicolumn{1}{c}{(0.0354)} & \multicolumn{1}{c}{(0.0351)} \\
    \multicolumn{1}{r}{} &       &       &       &       &       &  \\
    \multicolumn{1}{l}{Observations} &       & \multicolumn{1}{c}{1268} & \multicolumn{1}{c}{1268} &       & \multicolumn{1}{c}{1065} & \multicolumn{1}{c}{1064} \\
    \multicolumn{1}{r}{} &       &       &       &       &       &  \\
    \midrule
    \multicolumn{1}{r}{} &       &       &       &       &       &  \\
    \multicolumn{1}{l}{Demographic Controls} &       & \multicolumn{1}{c}{N} & \multicolumn{1}{c}{Y} &       & \multicolumn{1}{c}{N} & \multicolumn{1}{c}{Y} \\
    \multicolumn{1}{r}{} &       &       &       &       &       &  \\
    \midrule
    \midrule
    \multicolumn{7}{p{41em}}{\small Notes: Panel A reports the difference in the share of pro-CPTPP respondents between the groups who read the pro-CPTPP message bundled with an aligned and misaligned abortion message and the following two groups: (i) respondents who read only the pro-CPTPP message; and (ii) respondents in the No-Message condition. Panel B reports the difference in the share of pro-CPTPP respondents between the groups who read the anti-CPTPP message bundled with an aligned and misaligned abortion message and the following two groups: (i) respondents who read only the anti-CPTPP message; and (ii) respondents in the No-Message condition. All specifications include a flexible cubic spline in calendar fielding date (days since the start of data collection), with interior knots at quintiles of the field-date distribution. Demographics include three age categories, and indicators for whether the respondent identifies as man, is white, is Christian,  is a Republican or a Democrat,  has a college degree, identifies as upper class or upper-middle class, and is unemployed.  Robust standard errors in parentheses. ***\(p<0.01\), **\(p<0.05\), *\(p<0.1\).} \\
    \end{tabular}%
\end{adjustbox}
\end{table}

\newpage

\begin{table}[h!] 
  \centering
  \caption{Spillovers and Backlash from Abortion to Trade (Ordered Logit)}\label{ordered}
\begin{adjustbox}{width=\textwidth}
    \begin{tabular}{p{13em}llllll}
    \toprule
    \multicolumn{1}{r}{} &       &       &       &       &       &  \\
    \multicolumn{1}{r}{} &       & \multicolumn{5}{c}{Differences in Predicted Latent} \\
    \multicolumn{1}{r}{} &       & \multicolumn{5}{c}{Support for the CPTPP} \\
    \multicolumn{1}{r}{} &       & \multicolumn{2}{c}{} &       & \multicolumn{2}{c}{} \\
    \midrule
    \multicolumn{1}{r}{} &       &       &       &       &       &  \\
    \multicolumn{1}{r}{} &       & \multicolumn{5}{c}{Panel A. Pro-CPTPP Message } \\
    \multicolumn{1}{r}{} &       &       &       &       &       &  \\
    \multicolumn{1}{r}{} &       & \multicolumn{2}{c}{Relative to} &       & \multicolumn{2}{c}{Relative to} \\
    \multicolumn{1}{r}{} &       & \multicolumn{2}{c}{Pro-CPTPP Only} &       & \multicolumn{2}{c}{No Message} \\
\cmidrule{3-7}    \multicolumn{1}{l}{Pro-CPTPP, Aligned Abortion (\(\tau^{\text{Pro}}_A\))} &       & \multicolumn{1}{c}{0.2300 } & \multicolumn{1}{c}{0.2360 } &       & \multicolumn{1}{c}{0.3880 } & \multicolumn{1}{c}{0.422*} \\
    \multicolumn{1}{r}{} &       & \multicolumn{1}{c}{(0.2130)} & \multicolumn{1}{c}{(0.2140)} &       & \multicolumn{1}{c}{(0.2470)} & \multicolumn{1}{c}{(0.2490)} \\
     \multicolumn{1}{l}{Pro-CPTPP, Misaligned Abortion (\(\tau^{\text{Pro}}_M\))} &       & \multicolumn{1}{c}{-0.832***} & \multicolumn{1}{c}{-0.871***} &       & \multicolumn{1}{c}{-0.674***} & \multicolumn{1}{c}{-0.693***} \\
    \multicolumn{1}{r}{} &       & \multicolumn{1}{c}{(0.1810)} & \multicolumn{1}{c}{(0.1860)} &       & \multicolumn{1}{c}{(0.2190)} & \multicolumn{1}{c}{(0.2270)} \\
    \multicolumn{1}{r}{} &       &       &       &       &       &  \\
    \multicolumn{1}{l}{Observations} &       & \multicolumn{1}{c}{1254 } & \multicolumn{1}{c}{1251 } &       & \multicolumn{1}{c}{1050 } & \multicolumn{1}{c}{1047 } \\
    \multicolumn{1}{r}{} &       &       &       &       &       &  \\
    \midrule
    \multicolumn{1}{r}{} &       &       &       &       &       &  \\
    \multicolumn{1}{r}{} &       & \multicolumn{5}{c}{Panel B. Anti-CPTPP Message } \\
    \multicolumn{1}{r}{} &       &       &       &       &       &  \\
    \multicolumn{1}{r}{} &       & \multicolumn{2}{c}{Relative to} &       & \multicolumn{2}{c}{Relative to} \\
    \multicolumn{1}{r}{} &       & \multicolumn{2}{c}{Anti-CPTPP Only} &       & \multicolumn{2}{c}{No Message} \\
\cmidrule{3-7}    \multicolumn{1}{l}{Anti-CPTPP, Aligned Abortion (\(\tau^{\text{Anti}}_A\))} &       & \multicolumn{1}{c}{-0.0254} & \multicolumn{1}{c}{0.00647} &       & \multicolumn{1}{c}{-2.271***} & \multicolumn{1}{c}{-2.370***} \\
    \multicolumn{1}{r}{} &       & \multicolumn{1}{c}{(0.1400)} & \multicolumn{1}{c}{(0.1470)} &       & \multicolumn{1}{c}{(0.2130)} & \multicolumn{1}{c}{(0.2290)} \\
    \multicolumn{1}{l}{Anti-CPTPP, Misaligned Abortion (\(\tau^{\text{Anti}}_M\))} &       & \multicolumn{1}{c}{0.825***} & \multicolumn{1}{c}{0.860***} &       & \multicolumn{1}{c}{-1.421***} & \multicolumn{1}{c}{-1.507***} \\
    \multicolumn{1}{r}{} &       & \multicolumn{1}{c}{(0.1440)} & \multicolumn{1}{c}{(0.1480)} &       & \multicolumn{1}{c}{(0.2150)} & \multicolumn{1}{c}{(0.2270)} \\
    \multicolumn{1}{r}{} &       &       &       &       &       &  \\
    \multicolumn{1}{l}{Observations} &       & \multicolumn{1}{c}{1268} & \multicolumn{1}{c}{1268} &       & \multicolumn{1}{c}{1065} & \multicolumn{1}{c}{1064} \\
    \multicolumn{1}{r}{} &       &       &       &       &       &  \\
    \midrule
    \multicolumn{1}{r}{} &       &       &       &       &       &  \\
    \multicolumn{1}{l}{Demographic Controls} &       & \multicolumn{1}{c}{N} & \multicolumn{1}{c}{Y} &       & \multicolumn{1}{c}{N} & \multicolumn{1}{c}{Y} \\
    \multicolumn{1}{r}{} &       &       &       &       &       &  \\
    \midrule
    \midrule
    \multicolumn{7}{p{41em}}{\small Notes: Panel A reports the difference in predicted latent support for the CPTPP between the groups who read the pro-CPTPP message bundled with an aligned and misaligned abortion message and the following two groups: (i) respondents who read only the pro-CPTPP message; and (ii) respondents in the No-Message condition. Panel B reports the difference in predicted latent support for the CPTP between the groups who read the anti-CPTPP message bundled with an aligned and misaligned abortion message and the following two groups: (i) respondents who read only the anti-CPTPP message; and (ii) respondents in the No-Message condition. Estimation is by ordered logit. Demographics include three age categories, and indicators for whether the respondent identifies as man, is white, is Christian,  is a Republican or a Democrat,  has a college degree, identifies as upper class or upper-middle class, and is unemployed. Robust standard errors in parentheses. *** \(p<0.01\), ** \(p<0.05\), *\(p<0.1\).} \\
    \end{tabular}%
\end{adjustbox}
\end{table}

\newpage
\begin{table}[htbp]
\begin{adjustbox}{width = \textwidth}
  \centering
  \caption{Spillovers and Backlash from Abortion to Trade: Independents}\label{independents}
    \begin{tabular}{p{11.915em}llllll}
    \toprule
    \multicolumn{1}{r}{} &       &       &       &       &       &  \\
    \multicolumn{1}{r}{} &       & \multicolumn{5}{c}{Differences in Share of} \\
    \multicolumn{1}{r}{} &       & \multicolumn{5}{c}{ Respondents Supporting the CPTPP} \\
    \multicolumn{1}{r}{} &       & \multicolumn{2}{c}{} &       & \multicolumn{2}{c}{} \\
    \midrule
    \multicolumn{1}{r}{} &       &       &       &       &       &  \\
    \multicolumn{1}{r}{} &       & \multicolumn{5}{c}{Panel A. Pro-Trade Message } \\
    \multicolumn{1}{r}{} &       &       &       &       &       &  \\
    \multicolumn{1}{r}{} &       & \multicolumn{2}{c}{Relative to} &       & \multicolumn{2}{c}{Relative to} \\
    \multicolumn{1}{r}{} &       & \multicolumn{2}{c}{Pro-CPTPP Only} &       & \multicolumn{2}{c}{No Message} \\
\cmidrule{3-7}    \multicolumn{1}{l}{Pro-CPTPP, Aligned Abortion} &       & \multicolumn{1}{c}{0.0298} & \multicolumn{1}{c}{0.0346} &       & \multicolumn{1}{c}{-0.0158} & \multicolumn{1}{c}{-0.0162} \\
    \multicolumn{1}{r}{} &       & \multicolumn{1}{c}{(0.0414)} & \multicolumn{1}{c}{(0.0412)} &       & \multicolumn{1}{c}{(0.0499)} & \multicolumn{1}{c}{(0.0497)} \\
    \multicolumn{1}{l}{Pro-CPTPP, Misaligned Abortion} &       & \multicolumn{1}{c}{-0.180***} & \multicolumn{1}{c}{-0.176***} &       & \multicolumn{1}{c}{-0.226***} & \multicolumn{1}{c}{-0.230***} \\
    \multicolumn{1}{r}{} &       & \multicolumn{1}{c}{(0.0532)} & \multicolumn{1}{c}{(0.0544)} &       & \multicolumn{1}{c}{(0.0600)} & \multicolumn{1}{c}{(0.0616)} \\
    \multicolumn{1}{r}{} &       &       &       &       &       &  \\
    \multicolumn{1}{l}{Observations} &       & \multicolumn{1}{c}{374 } & \multicolumn{1}{c}{373 } &       & \multicolumn{1}{c}{292 } & \multicolumn{1}{c}{292 } \\
    \multicolumn{1}{r}{} &       &       &       &       &       &  \\
    \multicolumn{1}{r}{} &       &       &       &       &       &  \\
    \multicolumn{1}{r}{} &       & \multicolumn{5}{c}{Panel B. Anti-Trade Message } \\
    \multicolumn{1}{r}{} &       &       &       &       &       &  \\
    \multicolumn{1}{r}{} &       & \multicolumn{2}{c}{Relative to} &       & \multicolumn{2}{c}{Relative to} \\
    \multicolumn{1}{r}{} &       & \multicolumn{2}{c}{Anti-CPTPP Only} &       & \multicolumn{2}{c}{No Message} \\
\cmidrule{3-7}    \multicolumn{1}{l}{Anti-CPTPP, Aligned Abortion} &       & \multicolumn{1}{c}{0.0058 } & \multicolumn{1}{c}{0.0191 } &       & \multicolumn{1}{c}{-0.547***} & \multicolumn{1}{c}{-0.525***} \\
    \multicolumn{1}{r}{} &       & \multicolumn{1}{c}{(0.0552)} & \multicolumn{1}{c}{(0.0549)} &       & \multicolumn{1}{c}{(0.0565)} & \multicolumn{1}{c}{(0.0591)} \\
    \multicolumn{1}{l}{Anti-CPTPP, Misaligned Abortion} &       & \multicolumn{1}{c}{0.196***} & \multicolumn{1}{c}{0.190***} &       & \multicolumn{1}{c}{-0.356***} & \multicolumn{1}{c}{-0.356***} \\
    \multicolumn{1}{r}{} &       & \multicolumn{1}{c}{(0.0594)} & \multicolumn{1}{c}{(0.0588)} &       & \multicolumn{1}{c}{(0.0606)} & \multicolumn{1}{c}{(0.0607)} \\
    \multicolumn{1}{r}{} &       &       &       &       &       &  \\
    \multicolumn{1}{l}{Observations} &       & \multicolumn{1}{c}{428} & \multicolumn{1}{c}{428} &       & \multicolumn{1}{c}{332} & \multicolumn{1}{c}{331} \\
    \multicolumn{1}{r}{} &       &       &       &       &       &  \\
    \midrule
    \multicolumn{1}{r}{} &       &       &       &       &       &  \\
    \multicolumn{1}{l}{Demographic Controls} &       & \multicolumn{1}{c}{N} & \multicolumn{1}{c}{Y} &       & \multicolumn{1}{c}{N} & \multicolumn{1}{c}{Y} \\
    \multicolumn{1}{r}{} &       &       &       &       &       &  \\
    \midrule
    \midrule
    \multicolumn{7}{p{40em}}{\small Notes: The sample is restricted to respondents identified as Independents. Panel A reports the difference in the share of pro-CPTPP respondents between the groups who read the pro-CPTPP message bundled with an aligned and misaligned abortion message and the following two groups: (i) respondents who read only the pro-CPTPP message, and (ii) respondents in the No-Message condition. Panel B reports the difference in the share of pro-CPTPP respondents between the groups who read the anti-CPTPP message bundled with an aligned and misaligned abortion message and the following two groups: (i) respondents who read only the anti-CPTPP message; and (ii) respondents in the No-Message condition. Demographic controls include three age categories, and indicators for whether the respondent identifies as a man, is white, is Christian,  is a Republican or a Democrat,  has a college degree, identifies as upper-class or upper-middle class, and is unemployed.  Robust standard errors are in parentheses. *** \(p<0.01\), ** \(p<0.05\), *\(p<0.1\).} \\
    \end{tabular}%
\end{adjustbox}
\end{table}%
\newpage

\begin{sidewaystable}[ph!]
  \centering
\begin{adjustbox}{width = .9\textwidth}
  \caption{Spillover Effect and Backlash from Abortion to Trade: Controlling for Party Cues}\label{partycues}
    \begin{tabular}{p{13em}lllllllllllll}
    \toprule
    \multicolumn{1}{r}{} &       &       &       &       &       &       &       &       &       &       &       &       &  \\
    \multicolumn{1}{r}{} &       & \multicolumn{12}{c}{Differences in Share of  Respondents Supporting the CPTPP} \\
    \multicolumn{1}{r}{} &       &       &       &       &       &       &       &       &       &       &       &       &  \\
    \multicolumn{1}{r}{} &       & \multicolumn{2}{c}{} &       & \multicolumn{2}{c}{} &       &       &       &       &       &       &  \\
    \multicolumn{1}{r}{} &       &       &       &       &       &       &       &       &       &       &       &       &  \\
    \multicolumn{1}{r}{} &       & \multicolumn{5}{c}{Panel A. Pro-CPTPP Message } &       &       & \multicolumn{5}{c}{Panel B. Anti-CPTPP Message } \\
    \multicolumn{1}{r}{} &       &       &       &       &       &       &       &       &       &       &       &       &  \\
\cmidrule{3-7}\cmidrule{10-14}    \multicolumn{1}{r}{} &       & \multicolumn{2}{c}{Relative to} &       & \multicolumn{2}{c}{Relative to} &       &       & \multicolumn{2}{c}{Relative to} &       & \multicolumn{2}{c}{Relative to} \\
    \multicolumn{1}{r}{} &       & \multicolumn{2}{c}{Pro-CPTPP Only} &       & \multicolumn{2}{c}{No Message} &       &       & \multicolumn{2}{c}{Anti-CPTPP Only} &       & \multicolumn{2}{c}{No Message} \\
\cmidrule{3-7}\cmidrule{10-14}    \multicolumn{1}{l}{Aligned Abortion} &       & \multicolumn{1}{c}{-0.00597} & \multicolumn{1}{c}{0.00754} &       & \multicolumn{1}{c}{0.0135} & \multicolumn{1}{c}{-0.043} &       &       & \multicolumn{1}{c}{0.0111} & \multicolumn{1}{c}{0.0363} &       & \multicolumn{1}{c}{-0.466***} & \multicolumn{1}{c}{-0.502***} \\
    \multicolumn{1}{r}{} &       & \multicolumn{1}{c}{(0.03)} & \multicolumn{1}{c}{(0.04)} &       & \multicolumn{1}{c}{(0.04)} & \multicolumn{1}{c}{(0.05)} &       &       & \multicolumn{1}{c}{(0.04)} & \multicolumn{1}{c}{(0.05)} &       & \multicolumn{1}{c}{(0.04)} & \multicolumn{1}{c}{(0.05)} \\
    \multicolumn{1}{l}{Misaligned Abortion} &       & \multicolumn{1}{c}{-0.159***} & \multicolumn{1}{c}{-0.142***} &       & \multicolumn{1}{c}{-0.139***} & \multicolumn{1}{c}{-0.197***} &       &       & \multicolumn{1}{c}{0.148***} & \multicolumn{1}{c}{0.154***} &       & \multicolumn{1}{c}{-0.329***} & \multicolumn{1}{c}{-0.384***} \\
    \multicolumn{1}{r}{} &       & \multicolumn{1}{c}{(0.04)} & \multicolumn{1}{c}{(0.05)} &       & \multicolumn{1}{c}{(0.04)} & \multicolumn{1}{c}{(0.06)} &       &       & \multicolumn{1}{c}{(0.04)} & \multicolumn{1}{c}{(0.05)} &       & \multicolumn{1}{c}{(0.04)} & \multicolumn{1}{c}{(0.06)} \\
    \multicolumn{1}{r}{} &       &       &       &       &       &       &       &       &       &       &       &       &  \\
    \multicolumn{1}{l}{Aligned Party} &       & \multicolumn{1}{c}{0.0438 } & \multicolumn{1}{c}{0.0253 } &       & \multicolumn{1}{c}{0.0438 } & \multicolumn{1}{c}{0.120*} &       &       & \multicolumn{1}{c}{-0.0524} & \multicolumn{1}{c}{-0.0805} &       & \multicolumn{1}{c}{-0.0524} & \multicolumn{1}{c}{0.00692} \\
    \multicolumn{1}{r}{} &       & \multicolumn{1}{c}{(0.0326)} & \multicolumn{1}{c}{(0.0485)} &       & \multicolumn{1}{c}{(0.0326)} & \multicolumn{1}{c}{(0.0614)} &       &       & \multicolumn{1}{c}{(0.0421)} & \multicolumn{1}{c}{(0.0641)} &       & \multicolumn{1}{c}{(0.0421)} & \multicolumn{1}{c}{(0.0668)} \\
    \multicolumn{1}{l}{Misaligned Party} &       & \multicolumn{1}{c}{0.0420 } & \multicolumn{1}{c}{0.0199 } &       & \multicolumn{1}{c}{0.042} & \multicolumn{1}{c}{0.118*} &       &       & \multicolumn{1}{c}{0.106**} & \multicolumn{1}{c}{0.0972 } &       & \multicolumn{1}{c}{0.106**} & \multicolumn{1}{c}{0.183***} \\
    \multicolumn{1}{r}{} &       & \multicolumn{1}{c}{(0.0405)} & \multicolumn{1}{c}{(0.0540)} &       & \multicolumn{1}{c}{-0.0405} & \multicolumn{1}{c}{(0.0657)} &       &       & \multicolumn{1}{c}{(0.0444)} & \multicolumn{1}{c}{(0.0652)} &       & \multicolumn{1}{c}{(0.0444)} & \multicolumn{1}{c}{(0.0674)} \\
    \multicolumn{1}{r}{} &       &       &       &       &       &       &       &       &       &       &       &       &  \\
    \multicolumn{1}{l}{Observations} &       & \multicolumn{1}{c}{1254 } & \multicolumn{1}{c}{1251 } &       & \multicolumn{1}{c}{1050 } & \multicolumn{1}{c}{1047 } &       &       & \multicolumn{1}{c}{1268} & \multicolumn{1}{c}{1268} &       & \multicolumn{1}{c}{1065} & \multicolumn{1}{c}{1064} \\
    \multicolumn{1}{r}{} &       &       &       &       &       &       &       &       &       &       &       &       &  \\
    \midrule
    \multicolumn{1}{r}{} &       &       &       &       &       &       &       &       &       &       &       &       &  \\
    \multicolumn{1}{r}{} &       & \multicolumn{1}{c}{N} & \multicolumn{1}{c}{Y} &       & \multicolumn{1}{c}{N} & \multicolumn{1}{c}{Y} &       &       & \multicolumn{1}{c}{N} & \multicolumn{1}{c}{Y} &       & \multicolumn{1}{c}{N} & \multicolumn{1}{c}{Y} \\
    \multicolumn{1}{r}{} &       &       &       &       &       &       &       &       &       &       &       &       &  \\
    \midrule
    \midrule
    \multicolumn{14}{p{57.245em}}{Notes: Panel A reports the difference in the share of pro-CPTPP respondents between the groups who read the pro-CPTPP message bundled with an aligned and misaligned abortion message and the following two groups: (i) respondents who read only the pro-CPTPP message; and (ii) respondents in the No-Message condition. Panel B reports the difference in the share of pro-CPTPP respondents between the groups who read the anti-CPTPP message bundled with an aligned and misaligned abortion message and the following two groups: (i) respondents who read only the anti-CPTPP message; and (ii) respondents in the No-Message condition. All regressions include controls for whether the trade message was received bundled with the abortion stance distinctive of the party of the respondent (pro-choice for the Democratic Party, pro-life for the Republican Party). Demographic controls include three age categories, and indicators for whether the respondent identifies as man, is white, is Christian,  is a Republican or a Democrat,  has a college degree, identifies as upper class or upper-middle class, and is unemployed.  Robust standard errors in parentheses. *** \(p<0.01\), ** \(p<0.05\), *\(p<0.1\).} \\
    \end{tabular}%
\end{adjustbox}
\end{sidewaystable}

\newpage
\begin{table}[htbp]
  \centering
  \caption{Test of Priming Effects}\label{priming}
    \begin{tabular}{p{9.25em}llllll}
    \toprule
    \multicolumn{1}{r}{} &       &       &       &       &       &  \\
    \multicolumn{1}{r}{} &       & \multicolumn{5}{c}{Differences in Share of} \\
    \multicolumn{1}{r}{} &       & \multicolumn{5}{c}{ Respondents Supporting the CPTPP} \\
    \multicolumn{1}{r}{} &       & \multicolumn{5}{c}{(Relative to No Message)} \\
    \multicolumn{1}{r}{} &       &       &       &       &       &  \\
    \multicolumn{1}{r}{} &       & \multicolumn{2}{c}{Pro-Choice Respondents} &       & \multicolumn{2}{c}{Pro Life Respondents} \\
    \midrule
    \multicolumn{1}{r}{} &       &       &       &       &       &  \\
    \multicolumn{1}{l}{Pro-Life Message} &       & \multicolumn{1}{c}{0.0239} & \multicolumn{1}{c}{0.0266} &       & \multicolumn{1}{c}{0.00278} & \multicolumn{1}{c}{-0.00557} \\
    \multicolumn{1}{l}{} &       & \multicolumn{1}{c}{(0.0370)} & \multicolumn{1}{c}{(0.0374)} &       & \multicolumn{1}{c}{(0.0748)} & \multicolumn{1}{c}{(0.0788)} \\
    \multicolumn{1}{l}{} &       &       &       &       &       &  \\
    \multicolumn{1}{l}{Observations} &       & \multicolumn{1}{c}{305} & \multicolumn{1}{c}{304} &       & \multicolumn{1}{c}{120} & \multicolumn{1}{c}{119} \\
    \multicolumn{1}{l}{Demographic Controls} &       & \multicolumn{1}{c}{N} & \multicolumn{1}{c}{Y} &       & \multicolumn{1}{c}{N} & \multicolumn{1}{c}{Y} \\
    \multicolumn{1}{r}{} &       &       &       &       &       &  \\
    \midrule
    \midrule
    \multicolumn{7}{p{34.75em}}{Notes: The dependent variable is an indicator variable equal to 1 if the respondent supports the US membership in the CPTPP. Coefficients are differences relative to participants in the treatment arm where no message is sent. Demographic controls include three age categories, and indicators for whether the respondent identifies as a man, is white, is Christian,  is a Republican or a Democrat,  has a college degree, identifies as upper-class or upper-middle class, and is unemployed.  Robust standard errors are in parentheses. Sample size in square brackets. *** \(p<0.01\), **\(p<0.05\), *\(p<0.1\).} \\
    \end{tabular}%
\end{table}%

\newpage

\section{Motivating Evidence: Data Sources and Construction}\label{motgraphs}

This appendix documents the sources and coding used to construct Figure~\ref{ads} and Table~\ref{correlations}. Both are descriptive and are not used for identification in the experimental sections.

\begin{figure}[h!]
\centering
\includegraphics[width=.75\linewidth]{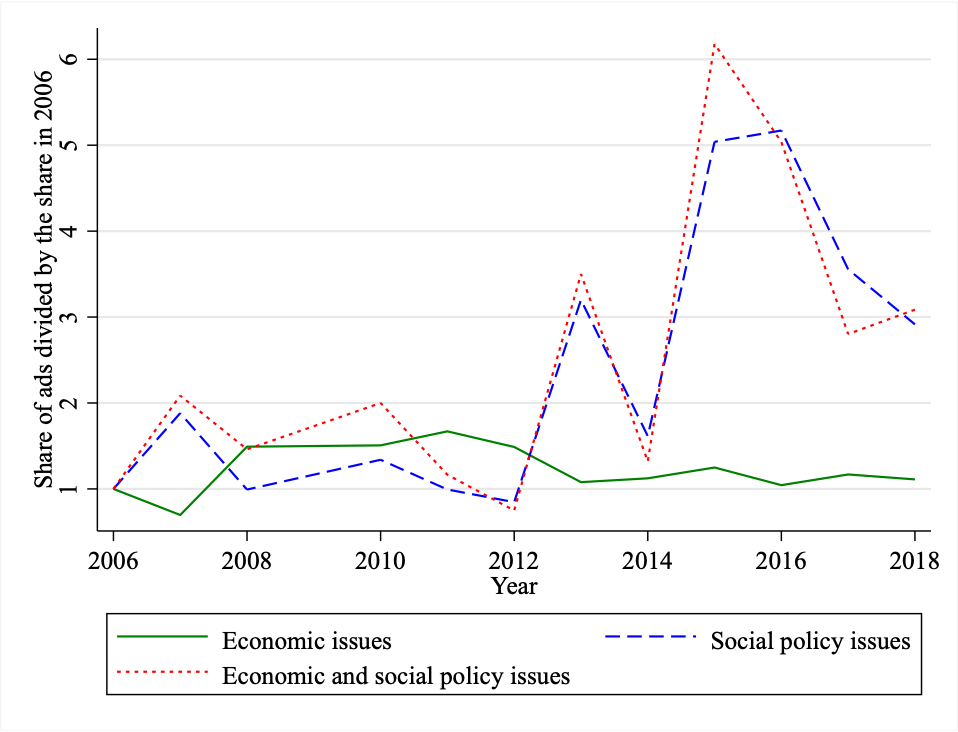}
    \caption{Change in Political Advertising by Topic: 2006--2018}\label{ads}
    \floatfoot{Notes. For each category (economic issues, social policy issues, or both), the figure reports the share of political television advertising that covers such topics in year $t$, divided by the same share in 2006. A political ad is classified as ``economic and social policy'' if it covers both domains. Source: Wesleyan Media Project.}
\end{figure}

\subsection{Figure~\ref{ads}: Political advertising topics (Wesleyan Media Project)}\label{motgraphs:ads}

Figure~\ref{ads} uses the Wesleyan Media Project (WMP) database of broadcast television political advertising. The unit of observation is an \emph{ad airing}. For each year $t$, let $n_t$ be the total number of tracked airings and let $n_t^{i}$ be the number of airings classified into topic category $i$.

\vspace{.5cm}

\paragraph{Topic categories.}
An airing is classified as \textbf{Economic} if it is tagged by WMP as covering at least one of:
\textit{Taxes; Deficit/Budget/Debt; Government Spending; Recession/Economic Stimulus; Minimum Wage; Farming; Business; Unions; Employment/Jobs; Poverty; Trade/Globalization; Housing/Sub-prime Mortgages; Economy (generic); Economic disparity/income inequality.}

An airing is classified as \textbf{Social policy} if it is tagged as covering at least one of:
\textit{Abortion; Homosexuality/Gay and Lesbian Rights; Moral/Family/Religious Values; Affirmative Action; Assisted Suicide/Euthanasia; Gun Control; Civil Rights/Racial Discrimination.}

An airing is classified as \textbf{Economic and social policy} if it satisfies both definitions.

\paragraph{Construction.}
For each category $i$, define the annual share $s_t^i\equiv n_t^i/n_t$. Figure~\ref{ads} plots the normalized series $s_t^i/s_{2006}^i$ for each category $i$ and each year $t\in\{2006,\ldots,2018\}$.

\subsection{Table~\ref{correlations}: Social--economic correlations}\label{motgraphs:correlations}

Table~\ref{correlations} uses Pew Research Center surveys fielded in 2004 and 2019, with the 2017 Political Typology survey used for pairs of variables not jointly available in the 2019 wave.\footnote{In the 2019 Pew wave, the abortion item is asked only to a subset of respondents. When abortion and a particular economic item are not jointly observed, the corresponding abortion--economic correlation uses the 2017 Political Typology survey, following the original draft.}
\vspace{.5cm}

\begin{table}[h!]
  \centering
\begin{adjustbox}{max width = 0.95\textwidth}
\caption{Growing Association of Social Policy and Economics}\label{correlations}
    \begin{tabular}{rrrrrrrrrrrr}
    \toprule
          &       &       &       &       &       &       &       &       &       &       &  \\
    \multicolumn{12}{c}{Pearson Correlation between Social Policy and Economic Policy Views} \\
          &       &       &       &       &       &       &       &       &       &       &  \\
    \midrule
          &       &       &       &       &       &       &       &       &       &       &  \\
          & \multicolumn{2}{c}{Gvt Efficiency} &       & \multicolumn{2}{c}{Gvt Regulation} &       & \multicolumn{2}{c}{Gvt Assist} &       & \multicolumn{2}{c}{Company Greed} \\
          & \multicolumn{1}{c}{2004} & \multicolumn{1}{c}{2019} &       & \multicolumn{1}{c}{2004} & \multicolumn{1}{c}{2019} &       & \multicolumn{1}{c}{2004} & \multicolumn{1}{c}{2019} &       & \multicolumn{1}{c}{2004} & \multicolumn{1}{c}{2019} \\
    \midrule
          &       &       &       &       &       &       &       &       &       &       &  \\
    \multicolumn{1}{l}{Gay Marriage} & \multicolumn{1}{c}{0.0581} & \multicolumn{1}{c}{0.2636} &       & \multicolumn{1}{c}{0.107} & \multicolumn{1}{c}{0.3357} &       & \multicolumn{1}{c}{0.1394} & \multicolumn{1}{c}{0.3164} &       & \multicolumn{1}{c}{0.0722} & \multicolumn{1}{c}{0.2945} \\
    \multicolumn{1}{l}{Abortion} & \multicolumn{1}{c}{0.0142} & \multicolumn{1}{c}{0.2298} &       & \multicolumn{1}{c}{0.052} & \multicolumn{1}{c}{0.3702} &       & \multicolumn{1}{c}{0.0936} & \multicolumn{1}{c}{0.2791} &       & \multicolumn{1}{c}{0.031} & \multicolumn{1}{c}{0.1783} \\
    \multicolumn{1}{l}{Discrimination} & \multicolumn{1}{c}{0.0389} & \multicolumn{1}{c}{0.2926} &       & \multicolumn{1}{c}{0.1024} & \multicolumn{1}{c}{0.3759} &       & \multicolumn{1}{c}{0.2372} & \multicolumn{1}{c}{0.4916} &       & \multicolumn{1}{c}{0.1612} & \multicolumn{1}{c}{0.342} \\
    \multicolumn{1}{l}{Immigration} & \multicolumn{1}{c}{0.164} & \multicolumn{1}{c}{0.2375} &       & \multicolumn{1}{c}{0.133} & \multicolumn{1}{c}{0.3595} &       & \multicolumn{1}{c}{0.0489} & \multicolumn{1}{c}{0.2896} &       & \multicolumn{1}{c}{-0.1278} & \multicolumn{1}{c}{0.2706} \\
          &       &       &       &       &       &       &       &       &       &       &  \\
    \midrule
    \midrule
          &       &       &       &       &       &       &       &       &       &       &  \\
\multicolumn{12}{p{45em}}{\small Notes: The table reports the Pearson correlation coefficient between views on four social policy issues (gay marriage, abortion, race discrimination, and immigration) and four economic policy issues (whether the government is wasteful, should regulate the economy, spend more to assist the poor, and whether big companies make too much profit) in 2004 and 2019. All variables are binary. For social policy issues, higher values denote more progressive views. For economic policy issues, higher values denote left-wing views. Source: \textit{Pew Research Center.}}
    \end{tabular}%
\end{adjustbox}
\end{table}

\paragraph{Economic policy indicators.}
Each economic item is coded as a binary indicator. For all items except Government Efficiency, the indicator equals 1 if the respondent chooses the first statement and 0 otherwise:
\begin{itemize}
\item Government regulation: ``necessary to protect the public interest'' (1) vs.\ ``usually does more harm than good'' (0).
\item Government assistance: ``should do more to help needy Americans...'' (1) vs.\ ``can't afford to do much more'' (0).
\item Company greed: ``corporations make too much profit'' (1) vs.\ ``fair and reasonable profit'' (0).
\item Government efficiency (reverse-coded): ``almost always wasteful and inefficient'' (0) vs.\ ``often does a better job than people give it credit for'' (1).
\end{itemize}

\paragraph{Social policy indicators.}
Abortion is coded as 1 for ``legal in all/most cases'' and 0 for ``illegal in most/all cases.'' Gay marriage is coded as 1 for ``very/somewhat good thing'' and 0 for ``somewhat/very bad thing.'' The other social items (discrimination, immigration) are coded as 1 for the more progressive statement and 0 for the more conservative statement, following the item wording in the survey.

\paragraph{Construction.}
For each year and each social--economic pair, Table~\ref{correlations} reports the Pearson correlation across respondents between the corresponding binary indicators.

\newpage
\section{Theory Appendix}\label{app:theory}

This appendix formalizes two benchmark mechanisms that organize interpretation of the experiments.
The first is a class of \emph{trust-based} Bayesian models in which moral alignment changes how informative or how aligned the sender is perceived to be. The second is an \emph{identity-threat} updating model in which moral conflict activates identity and induces motivated distancing.

% ==========================================================
% A. TRUST-BASED BENCHMARK
% ==========================================================
\subsection{Trust-based spillovers and backlash}\label{trustbased}

\paragraph{Environment.}
A receiver $R$ faces uncertainty about whether an economic policy has good consequences for her: $\omega^R\in\{0,1\}$.
A source $S$ takes an economic policy stance $y^S\in\{0,1\}$ (support/oppose). The receiver also observes (or fails to observe) the source's moral type $\theta\in\{A,M\}$: aligned ($A$) or misaligned ($M$) with the receiver on a salient moral issue. The receiver's \emph{mental model} is a joint distribution $\mathbb{P}$ over $(y^S,\omega^R,\theta)$ with full support. Suppose that \(\theta\) and \(\omega^R\) are independent.

Define the (policy-support) posterior as
\[
p(\theta)\equiv \mathbb{P}(\omega^R=1\mid y^S=1,\theta).
\]
Let the ``unknown type'' benchmark be
\[
p(U)\equiv \mathbb{P}(\omega^R=1\mid y^S=1),
\]
which corresponds to the situation where the receiver observes the economic stance but not the moral type.

\paragraph{Two trust channels.}
Both channels below imply that a misaligned sender is \emph{less persuasive} than an aligned sender. Let $\omega^S\in\{0,1\}$ denote whether the policy aligns with the sender objectives.

\emph{(i) Competence-based trust.} The interests of the parties are always aligned, \(\omega^S=\omega^R\), and type \(\theta\) affects the informativeness of $y^S$ about $\omega^R$. Formally, define the likelihood ratio
\[
\mathcal{I}(\theta)\equiv
\frac{\mathbb{P}(y^S=1\mid \omega^R=1,\theta)}{\mathbb{P}(y^S=1\mid \omega^R=0,\theta)}.
\]
Competence-based trust means $\mathcal{I}(A)>\mathcal{I}(M)\ge 1$, where the lower bound to competence is set to \(\mathcal{I}(\theta)=1\), so that in the worst case the senders' stance is uncorrelated with her objectives.

\emph{(ii) Preference-based trust.} Moral type affects perceived goal alignment between sender and receiver. Suppose $S$ supports iff $\omega^S=1$.
Assume \(\mathbb{P}(\omega^S=\omega^R\mid \omega^R, \theta)=\mathbb{P}(\omega^S=\omega^R\mid \theta)\) and let
\[
\mathcal{G}(\theta)\equiv \mathbb{P}(\omega^S=\omega^R\mid \theta)
\]
denote perceived goal alignment. Preference-based trust means $\mathcal{G}(A)>\mathcal{G}(M)$.

\paragraph{Spillovers and backlash.}
Both models imply \emph{disagreement spillovers}; preference-based trust can also produce backlash. Proposition \ref{prop0} summarizes why the empirical \emph{absence} of agreement spillovers is informative: in the canonical trust benchmark, $p(U)$ is a mixture of $p(A)$ and $p(M)$, so large negative spillovers mechanically tend to come with positive ones unless additional forces break the symmetry.

\begin{proposition}[Trust-based spillovers]\label{prop0}
Suppose either (i) competence-based trust or (ii) preference-based trust. Then:
\begin{enumerate}[(i)]
\item \textbf{Disagreement spillovers:} $p(M)<p(U)$.
\item \textbf{Agreement spillovers:} $p(A)>p(U)$.
\end{enumerate}
In addition, under preference-based trust, if $\mathcal{G}(M)$ is sufficiently low, the receiver can exhibit \textbf{backlash} relative to the prior, $p(M)<\mathbb{P}(\omega^R=1)$.
\end{proposition}

\begin{proof}[Proof of Proposition \ref{prop0}]
Write $\pi_0\equiv \mathbb{P}(\omega^R=1)$ for the receiver's prior. Throughout, full support implies
$\mathbb{P}(\theta=A\mid y^S=1)\in(0,1)$ and $\mathbb{P}(\theta=M\mid y^S=1)\in(0,1)$. By the law of total probability,
\[
p(U)=\mathbb{P}(\omega^R=1\mid y^S=1)
=\sum_{\theta\in\{A,M\}}\mathbb{P}(\omega^R=1\mid y^S=1,\theta)\mathbb{P}(\theta\mid y^S=1)
=\lambda\,p(A)+(1-\lambda)\,p(M),
\]
where $\lambda\equiv \mathbb{P}(\theta=A\mid y^S=1)\in(0,1)$.

\emph{(i) Competence-based trust.}
Let $\pi_0\equiv \mathbb{P}(\omega^R=1)$ denote the prior.
Bayes' rule gives posterior odds:
\[
\frac{p(\theta)}{1-p(\theta)}
=
\frac{\mathbb{P}(y^S=1\mid \omega^R=1,\theta)}{\mathbb{P}(y^S=1\mid \omega^R=0,\theta)}
\cdot
\frac{\mathbb{P}(\omega^R=1\mid \theta)}{\mathbb{P}(\omega^R=0\mid \theta)}
=
\mathcal{I}(\theta)\cdot \frac{\pi_0}{1-\pi_0},
\]
so
$\mathcal{I}(A)>\mathcal{I}(M)$ implies $p(A)>p(M)$.

\emph{(ii) Preference-based trust.}
Assume $S$ supports iff $\omega^S=1$ and that (conditional on $\theta$) the sender's and receiver's economic states coincide with probability $\mathcal{G}(\theta)$ and differ otherwise. Then, observing $y^S=1$ is equivalent to observing $\omega^S=1$, and
\[
p(\theta)
=\frac{\mathbb{P}(\omega^S=1\mid \omega^R=1,\theta)\mathbb{P}(\omega^R=1)}
{\mathbb{P}(\omega^S=1\mid \omega^R=1,\theta)\mathbb{P}(\omega^R=1)+\mathbb{P}(\omega^S=1\mid \omega^R=0,\theta)\mathbb{P}(\omega^R=0)}.
\]
Because $\mathbb{P}(\omega^S=1\mid \omega^R=1,\theta)=\mathcal{G}(\theta)$ and
$\mathbb{P}(\omega^S=1\mid \omega^R=0,\theta)=1-\mathcal{G}(\theta)$, we obtain
\begin{equation}\label{eq:p_theta_G}
p(\theta)=\frac{\mathcal{G}(\theta)\,\pi_0}{\mathcal{G}(\theta)\,\pi_0+(1-\mathcal{G}(\theta))(1-\pi_0)}.
\end{equation}
The expression is strictly increasing in $\mathcal{G}(\theta)$, so $\mathcal{G}(A)>\mathcal{G}(M)$ implies $p(A)>p(M)$.

But $p(U)=\lambda p(A)+(1-\lambda)p(M)$ with $\lambda\in(0,1)$ by full support. Hence $p(A)>p(M)$ implies
\[
p(M)<p(U)<p(A).
\]

Finally, under preference-based trust, using \ref{eq:p_theta_G} one can compare $p(\theta)$ to the prior $\pi_0$:
\[
p(\theta)<\pi_0
\iff
\frac{\mathcal{G}(\theta)\,\pi_0}{\mathcal{G}(\theta)\,\pi_0+(1-\mathcal{G}(\theta))(1-\pi_0)}<\pi_0
\iff
\mathcal{G}(\theta)<\tfrac12,
\]
where the last equivalence uses $\pi_0\in(0,1)$. Hence, if $\mathcal{G}(M)<1/2$, then $p(M)<\pi_0$ (backlash).

Under competence-based trust, the posterior-odds formula gives
\[
\frac{p(\theta)}{1-p(\theta)}=\mathcal{I}(\theta)\cdot\frac{\pi_0}{1-\pi_0}.
\]
If $\mathcal{I}(\theta)\ge 1$, then posterior odds are weakly higher than prior odds, so $p(\theta)\ge \pi_0$; thus backlash cannot occur.
\end{proof}

\subsection{Identity threat and motivated distancing}\label{identitymodel}

\paragraph{Environment.}
Fix an economic recommendation \(r\in\{0,1\}\), where \(r=1\) denotes support for the policy and \(r=0\) denotes opposition to it. Let \(a\in\{0,1\}\) denote the receiver's economic action, and let \(\omega^R\in\{0,1\}\) be the welfare-relevant state for the receiver, where action \(a=\omega^R\) is the action that is actually in her interest.

After observing the economic recommendation alone, the receiver forms a baseline posterior
\[
\pi(a)\equiv \mathbb{P}(\omega^R=a\mid y^S=r), \qquad a\in\{0,1\},
\]
with full support. Throughout this subsection, \(r\) is fixed, so the dependence of \(\pi\) on \(r\) is suppressed to lighten notation. Define
\[
p(U)\equiv \pi(r)\in(0,1)
\]
as the posterior probability that the recommended action is correct when the receiver observes only the economic recommendation, and let
\[
p_0\equiv \mathbb{P}(\omega^R=r)\in(0,1)
\]
denote the corresponding pre-message prior.

To connect the model to observed behavior, let \(Z_r\in\{0,1\}\) denote an indicator that the receiver chooses the \emph{recommended} action \(r\). Assume that, conditional on any information set \(\mathcal I\), the probability of following the recommendation weakly increases in the posterior probability that the recommendation is correct, so that
\(\mathbb{P}(Z_r=1\mid \mathcal I)\) is weakly increasing in \(\mathbb{P}(\omega^R=r\mid \mathcal I)\). When \(r=1\), \(Z_r\) coincides with policy support, and when \(r=0\), \(Z_r\) coincides with policy opposition. Thus the model's posterior \(p(\theta,c)\) maps directly into willingness to \emph{follow the recommendation}, while the sign of the effect on raw policy support depends on whether the recommendation is pro or anti.

\paragraph{Reference beliefs and identity activation.}
Let \(I\) and \(O\) denote the receiver's moral in-group and out-group. For each \(j\in\{I,O\}\), each economic action \(a\in\{0,1\}\), each moral-type realization \(\theta\in\{A,M\}\), and each source structure \(c\in\{0,1\}\), let
\[
q_j(a\mid \theta,c)\in(0,1), \qquad q_j(0\mid \theta,c)+q_j(1\mid \theta,c)=1.
\]
Here \(A\) denotes moral alignment and \(M\) moral misalignment; \(c=1\) means that the moral and economic statements are attributed to the \emph{same} source, while \(c=0\) means that they are attributed to \emph{different} sources. The object \(q_j(a\mid \theta,c)\) is the receiver's \emph{reference belief} that a typical member of group \(j\) would endorse action \(a\), after processing the message structure \((\theta,c)\). Equivalently, it measures how characteristic action \(a\) is of group \(j\) in the receiver's mental representation.

Let \(\eta(\theta)\ge 0\) denote the intensity with which the moral cue activates identity. The binary-action version of the identity-distorted updating rule is
\begin{equation}\label{eq:identity_app}
\tilde{\pi}_{\theta,c}(a)
=
\frac{
\pi(a)\left(\dfrac{q_I(a\mid \theta,c)}{q_O(a\mid \theta,c)}\right)^{\eta(\theta)}
}{
\sum_{a'\in\{0,1\}}
\pi(a')
\left(\dfrac{q_I(a'\mid \theta,c)}{q_O(a'\mid \theta,c)}\right)^{\eta(\theta)}
},
\qquad a\in\{0,1\}.
\end{equation}
This is the binary specialization of the identity-distorted updating rule used in \citet{BGT21,GT23}. In the present application, the novel ingredient is that the reference-belief objects \(q_j(\cdot\mid \theta,c)\) are allowed to depend on whether the moral cue identifies the speaker as aligned or misaligned and on whether the moral and economic statements are attached to the same source.

For the recommended action \(r\), define
\[
p(\theta,c)\equiv \tilde{\pi}_{\theta,c}(r).
\]
This is the posterior probability that the recommendation is correct after the receiver processes both the economic recommendation and the morally diagnostic cue.

Finally, define the \emph{net group-typing} of the recommended action by
\begin{equation}\label{eq:dr_app}
d_r(\theta,c)
\equiv
\log\!\left(
\frac{q_I(r\mid \theta,c)/q_O(r\mid \theta,c)}
     {q_I(1-r\mid \theta,c)/q_O(1-r\mid \theta,c)}
\right).
\end{equation}
The sign of \(d_r(\theta,c)\) summarizes whether the recommended action \(r\) is perceived as relatively more characteristic of the in-group or the out-group. In particular, \(d_r(\theta,c)<0\) means that the recommended action is relatively more out-group-typed than the alternative.

\begin{lemma}[Binary log-odds representation]\label{lem:id_logit}
Let \(\Lambda(x)\equiv (1+e^{-x})^{-1}\) denote the logistic cdf. Under \ref{eq:identity_app}, for every \((\theta,c)\in\{A,M\}\times\{0,1\}\),
\begin{equation}\label{eq:logit_rep_app}
p(\theta,c)
=
\Lambda\!\left(\operatorname{logit} p(U)+\eta(\theta)\,d_r(\theta,c)\right).
\end{equation}
Equivalently,
\[
\operatorname{logit} p(\theta,c)
=
\operatorname{logit} p(U)+\eta(\theta)\,d_r(\theta,c).
\]
\end{lemma}

\begin{proof}[Proof of Lemma \ref{lem:id_logit}]
By \ref{eq:identity_app},
\[
\frac{p(\theta,c)}{1-p(\theta,c)}
=
\frac{\tilde{\pi}_{\theta,c}(r)}{\tilde{\pi}_{\theta,c}(1-r)}
=
\frac{\pi(r)}{\pi(1-r)}
\cdot
\frac{\left(\dfrac{q_I(r\mid \theta,c)}{q_O(r\mid \theta,c)}\right)^{\eta(\theta)}}
{\left(\dfrac{q_I(1-r\mid \theta,c)}{q_O(1-r\mid \theta,c)}\right)^{\eta(\theta)}}.
\]
Since \(p(U)=\pi(r)\) and \(\pi(1-r)=1-p(U)\), this becomes
\[
\frac{p(\theta,c)}{1-p(\theta,c)}
=
\frac{p(U)}{1-p(U)}
\exp\!\big(\eta(\theta)\,d_r(\theta,c)\big),
\]
where \(d_r(\theta,c)\) is defined in \ref{eq:dr_app}. Taking logs yields
\[
\operatorname{logit} p(\theta,c)
=
\operatorname{logit} p(U)+\eta(\theta)\,d_r(\theta,c).
\]
Applying \(\Lambda\) to both sides gives \ref{eq:logit_rep_app}.
\end{proof}

\begin{assumption}[Conflict-driven activation]\label{ass:activation}
\[
\eta(A)=0<\eta(M).
\]
\end{assumption}

\begin{assumption}[Misaligned same-source outgroup typing]\label{ass:outgroup}
\[
d_r(M,1)<0.
\]
\end{assumption}

\begin{assumption}[Attenuation under source separation]\label{ass:attenuation}
\[
|d_r(M,0)|<|d_r(M,1)|.
\]
\end{assumption}

Assumption \ref{ass:activation} is the sharp conflict-driven version of the model: moral disagreement activates identity, while moral agreement does not. Assumption \ref{ass:outgroup} says that when a morally misaligned speaker makes the economic recommendation and the two statements come from the same source, the recommended action becomes relatively more out-group-typed. Assumption \ref{ass:attenuation} formalizes the common-source requirement: when the moral and economic statements are split across sources, the recommended action is more weakly associated with groups.

\begin{proposition}[Identity-threat implications]\label{prop1}
Fix a recommendation \(r\in\{0,1\}\). Under \ref{eq:identity_app} and
Assumptions \ref{ass:activation}--\ref{ass:attenuation}:

\begin{enumerate}[(i)]
\item \textbf{Disagreement spillover:}
\[
p(M,1)<p(U).
\]

\item \textbf{No agreement spillover in the sharp version:}
\[
p(A,c)=p(U)\qquad \text{for each } c\in\{0,1\}.
\]

\item \textbf{Attenuation under source separation:}
\[
\big|\operatorname{logit} p(M,0)-\operatorname{logit} p(U)\big|
<
\big|\operatorname{logit} p(M,1)-\operatorname{logit} p(U)\big|.
\]
Moreover,
\[
p(M,1)<p(M,0).
\]
The sign of \(p(M,0)-p(U)\) is unrestricted.

\item \textbf{Closeness to the economic-only benchmark:}
\[
|p(M,0)-p(U)|\le \frac{\eta(M)}{4}\,|d_r(M,0)|.
\]
Hence, if \(|d_r(M,0)|\) is small, then \(p(M,0)\) is close to \(p(U)\).

\item \textbf{Backlash threshold:} if \(p(U)>p_0\), define
\[
\eta^\star
\equiv
\frac{\operatorname{logit} p(U)-\operatorname{logit} p_0}{-d_r(M,1)}>0.
\]
Then
\[
p(M,1)<p_0
\qquad\Longleftrightarrow\qquad
\eta(M)>\eta^\star.
\]
\end{enumerate}
\end{proposition}

\begin{proof}[Proof of Proposition \ref{prop1}]
By Lemma \ref{lem:id_logit},
\[
\operatorname{logit} p(\theta,c)
=
\operatorname{logit} p(U)+\eta(\theta)\,d_r(\theta,c).
\]

\emph{Part (i).}
By Assumption \ref{ass:activation}, \(\eta(M)>0\). By Assumption \ref{ass:outgroup},
\(d_r(M,1)<0\). Hence
\[
\operatorname{logit} p(M,1)
=
\operatorname{logit} p(U)+\eta(M)d_r(M,1)
<
\operatorname{logit} p(U),
\]
so \(p(M,1)<p(U)\).

\emph{Part (ii).}
By Assumption \ref{ass:activation}, \(\eta(A)=0\). Therefore
\[
\operatorname{logit} p(A,c)=\operatorname{logit} p(U)
\]
for each \(c\in\{0,1\}\), hence \(p(A,c)=p(U)\).

\emph{Part (iii).}
Again by Lemma \ref{lem:id_logit},
\[
\operatorname{logit} p(M,0)-\operatorname{logit} p(U)
=
\eta(M)d_r(M,0),
\]
and
\[
\operatorname{logit} p(M,1)-\operatorname{logit} p(U)
=
\eta(M)d_r(M,1).
\]
Since \(\eta(M)>0\), Assumption \ref{ass:attenuation} implies
\[
\big|\operatorname{logit} p(M,0)-\operatorname{logit} p(U)\big|
<
\big|\operatorname{logit} p(M,1)-\operatorname{logit} p(U)\big|.
\]

To compare \(p(M,1)\) and \(p(M,0)\), write \(d_1\equiv d_r(M,1)\) and
\(d_0\equiv d_r(M,0)\). Assumption \ref{ass:outgroup} gives \(d_1<0\), and
Assumption \ref{ass:attenuation} gives \(|d_0|<|d_1|=-d_1\). Therefore
\[
d_1<d_0.
\]
Multiplying by \(\eta(M)>0\) and adding \(\operatorname{logit} p(U)\),
\[
\operatorname{logit} p(M,1)<\operatorname{logit} p(M,0),
\]
so \(p(M,1)<p(M,0)\). No sign restriction on \(d_0\) is imposed, so the sign of
\(p(M,0)-p(U)\) is unrestricted.

\emph{Part (iv).}
Let \(\Lambda(x)\equiv (1+e^{-x})^{-1}\). Then
\[
p(M,0)=\Lambda\!\left(\operatorname{logit} p(U)+\eta(M)d_r(M,0)\right),
\qquad
p(U)=\Lambda\!\left(\operatorname{logit} p(U)\right).
\]
By the mean value theorem, for some \(\xi\) between
\(\operatorname{logit} p(U)\) and \(\operatorname{logit} p(U)+\eta(M)d_r(M,0)\),
\[
p(M,0)-p(U)=\Lambda'(\xi)\,\eta(M)d_r(M,0).
\]
Since \(\Lambda'(x)=\Lambda(x)(1-\Lambda(x))\le 1/4\) for all \(x\),
\[
|p(M,0)-p(U)|\le \frac{\eta(M)}{4}|d_r(M,0)|.
\]

\emph{Part (v).}
If \(p(U)>p_0\), then \(\operatorname{logit} p(U)-\operatorname{logit} p_0>0\).
Using Lemma \ref{lem:id_logit},
\[
p(M,1)<p_0
\iff
\operatorname{logit} p(U)+\eta(M)d_r(M,1)<\operatorname{logit} p_0.
\]
Since \(d_r(M,1)<0\), this is equivalent to
\[
\eta(M)>
\frac{\operatorname{logit} p(U)-\operatorname{logit} p_0}{-d_r(M,1)}
=
\eta^\star.
\]
\end{proof}

\begin{remark}[Weak agreement spillovers]\label{rem:weakagree}
The sharp conflict-driven version sets \(\eta(A)=0\). A weaker version allows \(\eta(A)>0\) but small. In that case, if \(d_r(A,1)\ge 0\), then
\[
p(A,1)\ge p(U),
\]
with strict inequality whenever \(d_r(A,1)>0\). The magnitude of the agreement spillover is small whenever the product \(\eta(A)\,d_r(A,1)\) is small.
\end{remark}

\end{document}